\documentclass[aps,prb,amsmath,amssymb,reprint,superscriptaddress,citeautoscript,floatfix]{revtex4-2}

\usepackage{graphicx}
\usepackage{comment}
\usepackage{dcolumn}
\usepackage{bm}
\usepackage{microtype}
\usepackage{mathptmx}
\usepackage[colorlinks,linkcolor=blue,citecolor=blue,urlcolor=blue]{hyperref}
\usepackage{ifthen}
\usepackage{xparse}
\usepackage{color}
\usepackage{xcolor}

\def\CC{\ensuremath{\text{C}_2}}
\def\CN{\ensuremath{\text{C}_{\text{N}}}}
\def\CB{\ensuremath{\text{C}_{\text{B}}}}
\def\CNCBCN{\ensuremath{\text{C}_2\text{C}_{\text{N}}}}
\def\CBCNCB{\ensuremath{\text{C}_2\text{C}_{\text{B}}}}
\def\CNON{\ensuremath{\text{C}_{\text{N}}\text{O}_{\text{N}}}}
\def\CBVN{\ensuremath{\text{C}_{\text{B}}V_{\text{N}}}}
\def\CNVB{\ensuremath{\text{C}_{\text{N}}V_{\text{B}}}}
\def\VBON{\ensuremath{V_{\text{B}}\text{O}_{\text{N}}}}
\def\ON{\ensuremath{\text{O}_{\text{N}}}}
\def\CBVB{\ensuremath{\text{C}_{\text{B}}V_{\text{B}}}}
\def\CBCNt{\ensuremath{\text{C}_{\text{B}}{(\text{C}_{\text{N}})}_3}}
\def\CNCBt{\ensuremath{\text{C}_{\text{N}}{(\text{C}_{\text{B}})}_3}}
\def\CFCN{\ensuremath{\text{C}_4\text{C}_{\text{N}}}}
\def\CFCB{\ensuremath{\text{C}_4\text{C}_{\text{B}}}}

\newboolean{draft}   
\setboolean{draft}{true}
\NewDocumentCommand\todo{m}%
{%
  {\color{blue} (TODO: #1)}%
}
\NewDocumentCommand\change{om}%
{%
  \ifthenelse{\boolean{draft}}{
    \IfNoValueTF{#1}{}%
    {%
      {\color{gray}[#1]}
    }%
    {\color{orange}#2}%
  }%
  {#2}%
}

\begin{document}

\title{Thermodynamics of carbon point defects in hexagonal boron nitride}

\author{Marek Maciaszek}
\email{marek.maciaszek@pw.edu.pl}
\affiliation{ Faculty of Physics, Warsaw University of Technology, Koszykowa 75, 00-662 Warsaw, Poland}
\affiliation{Center for Physical Sciences and Technology (FTMC), Vilnius LT-10257, Lithuania}

\author{Lukas Razinkovas}
\affiliation{Center for Physical Sciences and Technology (FTMC), Vilnius LT-10257, Lithuania}

\author{Audrius Alkauskas}
\email{audrius.alkauskas@ftmc.lt}
\affiliation{Center for Physical Sciences and Technology (FTMC), Vilnius LT-10257, Lithuania}
\affiliation{Department of Physics, Kaunas University of Technology, Kaunas LT-51368, Lithuania}

\date{\today}

%%%%%%%%%%%%%%%%%%%%%%%%%%%%%%%%%%%%%%%%%%%%%%%%%%%%%%%%%%%%%%%%%%%%%%
%%%%%%%%%%%%%%%%%%%%%%%%% ABSTRACT %%%%%%%%%%%%%%%%%%%%%%%%%%%%%%%%%%%
%%%%%%%%%%%%%%%%%%%%%%%%%%%%%%%%%%%%%%%%%%%%%%%%%%%%%%%%%%%%%%%%%%%%%%

\begin{abstract}
  We present a first-principles computational study of the
  thermodynamics of carbon defects in hexagonal boron nitride
  (hBN). The defects considered are carbon monomers, dimers, trimers,
  and larger carbon clusters, as well as complexes of carbon with
  vacancies, antisites, and substitutional oxygen. Our calculations
  show that monomers ($\CB$, $\CN$), dimers, trimers, and $\CNON$ pairs 
  are the most prevalent species under most growth conditions. Compared 
  to these defects, larger carbon clusters, as well as complexes of 
  carbon with vacancies and antisites, occur at much smaller concentrations. 
  Our results are discussed in view of the relevance of carbon defects in 
  single-photon emission in hBN.
\end{abstract}

\maketitle

%%%%%%%%%%%%%%%%%%%%%%%%%%%%%%%%%%%%%%%%%%%%%%%%%%%%%%%%%%%%%%%%%%%%%%
%%%%%%%%%%%%%%%%%%%%% INTRODUCTION %%%%%%%%%%%%%%%%%%%%%%%%%%%%%%%%%%%
%%%%%%%%%%%%%%%%%%%%%%%%%%%%%%%%%%%%%%%%%%%%%%%%%%%%%%%%%%%%%%%%%%%%%%

\section{Introduction\label{sec:intro}}

{
\raggedbottom

Over the past two decades, hexagonal boron nitride (hBN) has gained
significant interest as both an insulator used in conjunction with
other layered two-dimensional materials~\cite{Duong2017} and as an
active optical medium itself~\cite{Gil2020}. The large bandgap of hBN,
6.08~eV~\cite{Cassabois2016}, makes it unique among the layered
compounds.  For all potential applications of hBN, whether as a
passive insulator or as an active optoelectronic material, the purity
of single crystals or epitaxial layers is crucial. Unfortunately, the
growth of ultra-pure material has been challenging~\cite{sun2018}.  In
part, this is because the defect chemistry of boron nitride is still
poorly understood.

Due to an intermediate position in the periodic table between boron
and nitrogen and a similar crystal structure, carbon 
plays an exceptional role in hBN.  It is known that carbon is often
present as an unintentional impurity in hBN grown by a variety of
different methods. Experiments on bulk hBN samples produced by
high-pressure high-temperature synthesis with Ba--BN as a
solvent~\cite{Taniguchi2007}, a frequently employed growth method,
showed the existence of carbon-rich domains~\cite{Onodera2019}. These
domains extended both along the $c$-axis as well as in the 2D layers,
and the density of carbon atoms in these domains exceeded
$10^{18}~\mbox{cm}^{-3}$~\cite{Onodera2019}, as indicated by
secondary-ion mass spectroscopy. The presence of carbon in hBN was
also revealed by direct imaging via the annular dark-field electron
microscopy~\cite{Krivanek2010} on samples exfoliated from bulk hBN.

The interest in point defects in hBN in general and carbon-related
defects in particular has recently increased due to the observation
of single-photon
emission in this material~\cite{Tran2016,Bourrellier2016,Grosso2017,Exarhos2017}. The
emission is believed to originate at point defects; for an overview of
this fast-developing field, we refer to
Ref.~\onlinecite{sajid2020a}. Compared to materials such as diamond or
silicon carbide, a two-dimensional material like hBN offers certain
advantages as a host of quantum emitters, e.g., the possibility of
high photon extraction efficiency. The reports about single-photon
emitters (SPEs) in hBN are numerous. Most of the emitters have structured
luminescence bands with with zero-phonon line (ZPL) energies in the range
\mbox{1.6--2.1}~eV~\mbox{\cite{Tran2016,Bourrellier2016,Grosso2017,Exarhos2017}}, accompanied by a clearly pronounced phonon sideband. In Ref.~\onlinecite{Mendelson2020},
Mendelson \textit{et al.} used various experimental techniques and
samples grown by several methods to show that single-photon emission
in this energy range is carbon-related (with most SPEs having ZPL
energies at about 2.1 eV). This means that emitters either directly
contain carbon atoms or carbon influences the formation of other
defects.  Several recent theoretical works suggested carbon-related
defects to explain emission in this energy range. In particular, Sajid
and Thygesen proposed the complex of a substitutional carbon with
a nitrogen vacancy, $\CBVN$, as one of the
candidates~\cite{sajid2020}.  Jara \emph{et al.} suggested that carbon
trimers $\text{C}_3$ ($\CBCNCB$ and $\CNCBCN$) have optical signatures
resembling some of the emitters in the \mbox{1.6--2.1~eV} range, also
in agreement with the involvement of carbon in single-photon
emission. The hypothesis about the $\CNCBCN$ trimer in particular has been recently
backed by \mbox{Li \emph{et al.}}~\cite{li2021}. Several calculated
optical properties of this defect matched the experimental
observations well.  Finally, Auburger and Gali
performed calculations of optical excitations energies and magnetic
resonance signatures of several carbon-containing defects. They
concluded that carbon on the boron site $\CB$ is the most likely
candidate to explain the emission~\cite{Auburger2021}.

Apart from quantum emission in the red and the infrared (IR) spectral
regions, in Ref.~\onlinecite{Bourrellier2016}, single-photon emission
in the near-ultraviolet (UV) range has been reported. The optical
signal of the emitter is nearly identical to the well-known 4.1~eV
luminescence line in hBN~\cite{Era1981,museur2008}. It was known for
some time that deliberate doping of hBN with carbon increases the
intensity of this line~\cite{Era1981,Uddin2017,Pelini2019}. Therefore,
the 4.1~eV line, frequently observed in nominally undoped
samples~\cite{Solozhenko2001}, has often been used as a proxy for
carbon in hBN. Historically, the 4.1~eV line was often attributed to
carbon on the nitrogen site \cite{Katzir1975}. However, it was recently shown that $\CN$
is not the origin of the 4.1~eV
luminescence~\cite{weston2018}. Instead, we have recently provided
theoretical evidence that the origin of this line is the carbon dimer
defect $\CC$~\cite{mackoit2019}. The $\CNON$ pair has also been
suggested to explain the 4.1~eV
luminescence~\cite{vokhmintsev2019}. However, a stoichiometric
Stone--Wales defect that does not involve carbon at all has also been
recently proposed~\cite{hamdi2020}. All these three defects are
expected to emit in the UV, but the debate regarding the exact
chemical nature of the 4.1~eV emitters seems to continue.\looseness=-1

It is clear that present evidence strongly suggests the importance of
carbon-related defects in optical emission in hBN, as carbon appears
to be an essential component of SPEs both in the red/IR and UV
regions. However, it can also be concluded that a universal consensus
regarding an assignment of a specific defect to a
particular luminescence line has not yet been
reached, despite intensive experimental and theoretical
investigations. This situation is certainly unsatisfactory, as it
hinders the development of hBN devices based on color centers. Most
of the theoretical investigations that suggested a certain defect
giving rise to a specific luminescence band analyzed one or at most a
few defect candidates. However, a broader picture of carbon defects in
hBN has been missing.

In this work, we present first-principles density functional theory
calculations of carbon in hBN.\ In particular, we study
single carbon substitutionals, carbon dimers, trimers,
non-nearest-neighbor carbon pairs, and larger carbon clusters, as
well as complexes of substitutional carbon with vacancies, antisites,
and substitutional oxygen. Our goal is to analyze the thermodynamics
of carbon-related species under various growth conditions. These
results provide a general picture of carbon-related defect chemistry
that will be helpful for interpreting existing
data and designing new experiments.

For the sake of completeness, we must state that a very different class of single-photon emitters
in hBN has been recently discovered. These emitters feature broad bands peaking at about
830 nm (${\sim}1.5$ eV). They are attributed to negatively-charged boron vacancies \cite{Gottscholl2020,Gotscholl2021}.
As of now, these are the only color centers in hBN where a consensus \cite{Ivady2020} regarding their chemical structure exists.

The paper is organized as follows. In Sec.~\ref{sec:th}, the
theoretical approach is outlined. In Sec.~\ref{sec:carb}, we discuss
defects analyzed in this work. Calculated formation energies and
equilibrium defect concentrations are analyzed in
Sec.~\ref{sec:formation}. The relevance of our calculations for
single-photon emission experiments is discussed in
Sec.~\ref{sec:expt}. In Sec.~\ref{sec:validity}, we analyze the range
of validity of the formation energy formalism for carbon-related
defects in hBN and present the results for solubility limits of carbon
in hBN. Sec.~\ref{sec:conc} concludes our paper.}

%%%%%%%%%%%%%%%%%%%%%%%%%%%%%%%%%%%%%%%%%%%%%%%%%%%%%%%%%%%%%%%%%%%%%%
%%%%%%%%%%%%%%%%%%%%% THEORETICAL METHODOLOGY %%%%%%%%%%%%%%%%%%%%%%%%
%%%%%%%%%%%%%%%%%%%%%%%%%%%%%%%%%%%%%%%%%%%%%%%%%%%%%%%%%%%%%%%%%%%%%%

\vspace{-0.5em}
\section{Theoretical methodology\label{sec:th}}
\vspace{-0.5em}

Our first-principles calculations were performed using the
Heyd--Scuseria--Ernzerhof ~\cite{HSE} hybrid functional based on
screened Fock exchange. The screening parameter
was~$0.2~\text{\AA}^{-1}$, and a mixing parameter, which indicates the
fraction of the screened Fock exchange ad-mixed to the local exchange,
was set to $a = 0.31$. The local exchange is described via a modified
Perdew--Burke--Ernerzof density functional~\cite{PBE}. The mixing
parameter was chosen to obtain a good description of the bandgap. The
plane-wave basis set with the kinetic energy cutoff 500~eV was used,
and the interaction of electrons and ions was described via the
projector-augmented wave approach~\cite{PAW}. Positions of ions were
relaxed until all forces fell below 0.01~eV/\AA.\ We used orthorhombic
supercells with 240 atoms to model defects. The lattice vectors of
supercells are
$5(\mathbf{a} + \mathbf{b}), 3(\mathbf{a} - \mathbf{b}), 2\mathbf{c}$,
where $\mathbf{a}$, $\mathbf{b}$, and $\mathbf{c}$ are primitive vectors
of the Bravais lattice.\ To take into account van der Waals interactions
between the hBN layers, we applied the Grimme-D3
scheme~\cite{Grimme2006}. Geometry relaxation was performed and total 
energies were computed including these van der Waals corrections.
A single $\Gamma$ point was used for the Brillouin zone
integration. Calculations have been performed using the Vienna
Ab-Initio Simulation Package (\textsc{Vasp})~\cite{vasp}. The
described methodology reproduces experimental parameters very
well. The calculated bandgap is 5.91~eV (experimental value
6.08~eV~\cite{Cassabois2016}), the formation enthalpy of the compound 
equals to $-2.89$~eV per formula unit (experimental value $-2.60$~eV \cite{Tomaszkiewicz2002}),
lattice constants are $a = 2.49$~\AA\ and
$c = 6.55$~\AA\ (experimental values 2.50~\AA\
and 6.65~\AA, respectively~\cite{Gu2007}).

Let us consider a point defect in a charge state $q$. Let
$n_{\text{C}}$, $n_{\text{B}}$, and $n_{\text{N}}$ be the number of
carbon, boron, and nitrogen atoms added to the hBN supercell when the
defect is created ($n_{\text{B}}$ and $n_{\text{N}}$ can be negative,
meaning that atoms have been removed). The formation energy of the
defect as a function of the electron chemical potential (or Fermi
level) $E_F$ is~\cite{Zhang1991,Freysoldt2014}
\begin{eqnarray}
  \Delta H_f(D,q)
  &
    =
  &
    E_{\mathrm{tot}}(D,q) - E_{\mathrm{bulk}}
    - n_{\text{C}} \mu_{\text{C}} - n_{\text{B}} \mu_{\text{B}} - n_{\text{N}}  \mu_{\text{N}}
    \nonumber
  \\
  && {}  + q(E_V + E_F ) + E_{\mathrm{corr}}.
\label{eq:formation}
\end{eqnarray}
Here, $q$ is the charge state of the defect, $E_{\mathrm{tot}}(D,q)$ is the total energy of the supercell
containing one defect, $E_{\mathrm{bulk}}$ is the total energy of the
defect-free supercell. $\mu_{\text{B}}$, $\mu_{\text{N}}$, and
$\mu_{\text{C}}$ are chemical potentials of boron, nitrogen, and
carbon, respectively. $E_V$ is the energy of the valence band maximum (VBM),
with respect to which $E_F$ is referenced. $E_{\mathrm{corr}}$ is a correction term
that accounts for spurious interactions of charged defects in periodic
supercells, and we use the form of the correction provided in
Ref.~\onlinecite{Freysoldt2009}. 
The value of the Fermi energy for
which charge states $q$ and $q'$ have equal formation energies is
called the charge-state transition level and is labelled
$(q/q')$~\cite{Freysoldt2014}. 

Atomic chemical potentials that appear in Eq.~\eqref{eq:formation}
reflect the chemical environment during growth. The equilibrium nature of
growth of hBN is ensured by the condition
$\mu_{\text{B}} + \mu_{\text{N}} = \mu_{\text{BN}}$, where
$\mu_{\text{BN}}$ is the total energy of bulk hBN per one formula
unit. Within the constraint of equilibrium growth, N-rich (B-poor)
conditions correspond to the value of $\mu_{\text{N}}=1/2 \mu_{N_2}$,
where $\mu_{N_2}$ is the energy of the $\text{N}_2$ molecule. Under
B-rich (N-poor) conditions
$\mu_{\text{B}}=\mu_{\text{B}_{\text{bulk}}}$, where
$\mu_{\text{B}_{\text{bulk}}}$ is the total energy of bulk boron per
one atom. The actual chemical potentials must always be within these
two limits. Therefore, we discuss formation energies for both N-rich
and N-poor conditions.  The chemical potential of carbon
$\mu_{\text{C}}$ is equal to the energy of carbon in bulk graphite.
This is the carbon-rich limit, which will be discussed in more detail 
in Secs.~\ref{sec:expt} and \ref{sec:validity}.

Apart from carbon, it is known that oxygen is another prevalent
impurity in hBN~\cite{Krivanek2010}. Therefore, we choose to model the
carbon chemistry in two situations: without and with oxygen. The
formation energy of oxygen-containing defects is calculated using an
expression similar to Eq.~\eqref{eq:formation}, where oxygen atoms are
accounted for via the oxygen chemical potential~$\mu_{\text{O}}$. We
consider the oxygen-rich limit. Due to a very low formation enthalpy
(high stability) of $\text{B}_2\text{O}_3$, the maximum chemical 
potential of oxygen is determined by the energy of O in $\text{B}_2\text{O}_3$ 
under both N-rich and N-poor conditions. The need to consider the formation
of $\text{B}_2\text{O}_3$ when determining the oxygen chemical
potential brings an additional complication. Despite the low
formation enthalpy of $\text{B}_2\text{O}_3$, its melting temperature
is ${\sim}750$~K, below typical growth temperatures of hBN ($T>1000$~K).\ In
this work, we do not consider the liquid phase and take the limiting
value of $\mu_{\text{O}}$ by considering solid
$\text{B}_2\text{O}_3$. This makes our calculations for O-containing
defects slightly more qualitative.

Once the formation energies of defects are known, their equilibrium
concentrations are determined using thermodynamical
simulations~\cite{Laks1992,Lany2007}. In brief, the concentration of
certain defect $D$ in a charge state $q$ at temperature $T$ is
determined via its formation energy:
\begin{equation}
N_{D,q} (T) = N_{D, \text{sites}}e^{-\Delta H_f(D,q)/k_B{T}},
\label{eq:ND}
\end{equation}
where $N_{D, \text{sites}}$ is the density of sites for the defect to
form (we use the terms ``density'' and ``concentration'' as synonyms). We express $N_{D, \text{sites}}$ via
$N_{D, \text{sites}}=g_DN_{\text{sites}}$, where $N_{\text{sites}}$ is
the density of atomic sites in hBN. This definition explains the physical
meaning of the factor $g_D$. 
In the literature, $g_D$ is sometimes called lattice multiplicity factor
or simply degeneracy factor. 
For example, $g_D=1/2$ 
for single substitutional defects in hBN; \mbox{$g_D=3/2$} for carbon dimers; $g_D=6$ for carbon tetramers
(\textit{cis} and \textit{trans} conformers have nearly the same
energies, see below); $g_D=1/2$ for closed-ring carbon hexamers;
etc. (for a description of different carbon defects, see
Sec.~\ref{sec:carb}). The position of the Fermi level $E_F$ is
determined by imposing the charge-neutrality condition
$\sum_{D,q} N_{D,q}q=0$~\cite{Laks1992,Lany2007}. The sum is carried 
over all defects included in the modelling. Due to a large bandgap of hBN, we
exclude free carriers from our simulations. We verified that including them does 
not affect the main results of the paper. Eq.~\eqref{eq:ND} for all considered 
defects and the charge-neutrality condition form a set of equations that are 
solved self-consistently.

The outlined methodology has several shortcomings that need to be
discussed. First, we state at the outset that we model only the
thermodynamics of carbon-related defects. In reality, kinetic effects
(e.g., diffusion) could lead to defect concentrations being different
than predicted by thermodynamics. Due to the complexity of the
problem, we abstain from discussing kinetic effects in this
work. Instead, our goal is to establish the limits set by equilibrium
thermodynamics. The second shortcoming is related to vibrational
contributions to free energies. More rigorously, one should determine
the formation enthalpy at finite temperatures. This means that
vibrational contribution to free energies should be
included in Eq.~\eqref{eq:formation}~\cite{Freysoldt2014}. In this
work, we approximate these energies by their zero-temperature
counterparts, as it is often done in defect simulations \cite{Freysoldt2014}.

\vspace{-0.5em}
\section{Carbon defects\label{sec:carb}}
\vspace{-0.5em}

\begin{figure*}
  \centering
\includegraphics[width=0.79\textwidth]{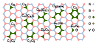}
\caption{A subset of carbon-related defects considered in this work:
  single substitutional defects $\CB$ and $\CN$, carbon dimer
  $\CC$, carbon trimers $\CNCBCN$ and $\CBCNCB$, carbon tetramer C$_4$ (both \emph{cis} and \emph{trans} configurations are shown),
  triangular star-like defects $\CNCBt$ and $\CBCNt$, hexamer $\text{C}_6$,
  the $\CBVN$ complex, and the $\CN\ON$ pair.\label{fig:defects}}
\vspace{-0.5em}
\end{figure*}

We include a large number of carbon defects in our analysis. These can
be grouped into several distinct sets. (i)~Carbon substitutionals and
their clusters: monomers $\CB$ and $\CN$, the dimer
$\CC$ (where the two carbon atoms are on adjacent sites), carbon pairs
C$_{\text{B}}$--C$_{\text{N}}$ (whereby the two carbon atoms are not
adjacent), trimers $\text{C}_3$ ($\CNCBCN$ and $\CBCNCB$),
triangular (star-like) defects $\CNCBt$ and $\CBCNt$, as well as
larger clusters, in particular $\text{C}_4$, $\text{C}_5$,
$\text{C}_6$, $\text{C}_8$, and $\text{C}_{10}$. (ii)~Complexes of
carbon substitutionals with vacancies: $\CBVB$ and
$\text{C}_{\text{N}}V_{\text{N}}$ (that can be thought of as
donor-acceptor pairs), $\CBVN$ (donor-donor pair), and $\CNVB$
(acceptor-acceptor pair). (iii)~Antisite-carbon pairs:
$\text{N}_{\text{B}}\text{C}_{\text{B}}$,
$\text{N}_{\text{B}}\text{C}_{\text{N}}$,
$\text{B}_{\text{N}}\text{C}_{\text{B}}$, and
$\text{B}_{\text{N}}\text{C}_{\text{N}}$. 
As it is clear from the list, we consider both first-neighbor and second-neighbor complexes.
Some of the studied defects are shown in Fig.~\ref{fig:defects}.

We exclude carbon interstitials and complexes of interstitials with
other defects from our analysis. A few of such defects were recently
considered in Ref.~\onlinecite{bhang2021}. They typically have very large
formation energies. In addition, due to a layered structure of hBN,
interstitials have very low diffusion barriers and thus should be
easily annealed~\cite{weston2018}.

Formation energies of charged defects depend on the Fermi level,
which is governed by the charge state of all point defects and 
impurities in the system. As a result, our thermodynamic modeling 
includes all other defects that might affect the position of the Fermi level. 
These defects are bare vacancies $V_{\text{B}}$ and $V_{\text{N}}$, as well as antisite
defects $\text{N}_{\text{B}}$ and $\text{B}_{\text{N}}$. These are the
intrinsic defects with lowest formation
energies~\cite{weston2018}. 

When oxygen is present, in addition to defects discussed above, we
also include the oxygen--carbon pair $\CNON$ (shown in Fig.~\ref{fig:defects}),
the substitutional oxygen $\ON$, and the boron vacancy--oxygen
complex $\VBON$ in our analysis. 

%%%%%%%%%%%%%%%%%%%%%%%%%%%%%%%%%%%%%%%%%%%%%%%%%%%%%%%%%%%%%%%%%%%%%%
%%%%%%%%%%%%%%%%%%%%%%%%%% FORMATION ENERGIES %%%%%%%%%%%%%%%%%%%%%%%%
%%%%%%%%%%%%%%%%%%%%%%%%%%%%%%%%%%%%%%%%%%%%%%%%%%%%%%%%%%%%%%%%%%%%%%

\section{Formation energies and densities \label{sec:formation}}

In this section, we explicitly discuss only defects that can occur at
concentrations $N_{D}>10^{14}~\mbox{cm}^{-3}$ for growth temperatures $T<2000$~K. 
The reasoning behind this criterion will be discussed in Sec.~\ref{sec:expt}. The data for all other carbon 
defects can be found in the Supplemental Material~\cite{supp}.  

\subsection{Formation energies\label{sec:energies}}

\begin{figure*}
\includegraphics[width=1\linewidth]{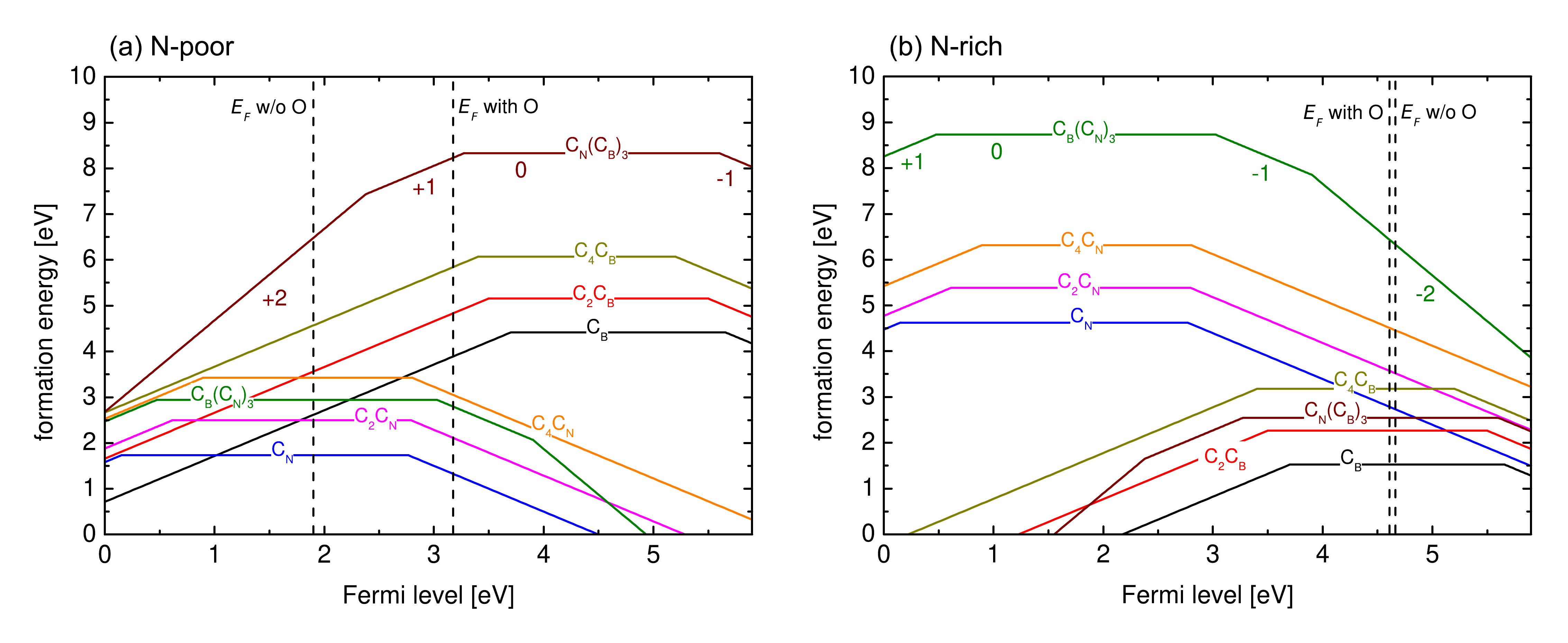}
\caption{Formation energies of odd carbon clusters as a function of
  the Fermi level. Defects considered are: single substitutional
  carbon atoms, carbon trimers, pentamers, as well as triangular
  (star-like) defects $\CNCBt$ and $\CBCNt$. (a)~N-poor conditions;
  (b)~N-rich conditions.  Vertical dashed lines show the values of our
  calculated Fermi level with and without oxygen.  Numbers close to
  the formation energy line indicate the charge state. For simplicity,
  the charge state is shown only for one defect. The charge states for
  other defects can be inferred from the slope of formation energy
  lines.\label{fig:C_comp_odd}}
\vspace{1em}
\end{figure*}

\textit{Odd carbon clusters}. Simple carbon substitutionals $\CB$ and
$\CN$ are carbon defects that have been investigated
previously~\cite{Huang2012,Berseneva2013,weston2018,mackoit2019,Auburger2021}. Since
single boron or nitrogen atom is replaced by an impurity carbon
atom, formation energies depend on the growth
conditions for these defects, i.e., N-rich (B-poor), B-rich (N-poor), or some
intermediate conditions. The results for ``extreme'' conditions are
shown in Fig.~\ref{fig:C_comp_odd} and are in agreement with previous
calculations, e.g., \mbox{Refs.~\onlinecite{weston2018,mackoit2019}}. In
addition to simple substitutional defects, in
Fig.~\ref{fig:C_comp_odd} we show the formation energies of carbon
clusters with an unequal number of carbons on boron or nitrogen sites:
carbon trimers, pentamers, as well as triangular
defects $\CNCBt$ and $\CBCNt$. 
In the formation energy plot, we depict only the most stable charge state for a given
Fermi level. This yields a piece-wise linear dependence of $\Delta H_f$ on $E_F$.
The actual charge state can be read-out from the slope of the line at a given $E_F$ \cite{Freysoldt2014}.

Vertical dashed
lines in Fig.~\ref{fig:C_comp_odd} indicate the values of the calculated
Fermi level. We choose $T=1600$~K as a representative temperature (see Sec.~\ref{sec:expt}), but
we find that $E_F$ depends quite weakly on $T$. The position
of the Fermi level is shown for the growth with and without
oxygen. Different situations will be discussed in
Sec.~\ref{sec:thermodynamics}.

We find that, in general, formation energies increase with the number
of carbon atoms in the cluster. Single substitutionals, as well as
trimers and pentamers, can have three charge states: $-1$, $0$, and
$+1$. From the chemical perspective, one can expect that, for example,
clusters with one additional $\CN$ unit are single acceptors. Therefore, the
existence of $0$ and $-1$ charge states is natural. However,
calculations show that a positive charge state is also stable
in all cases. A similar conclusion holds for
clusters with one more carbon atom on the boron site, where we also find
three charge states. In the case of
star-like defects, we also find higher charge states: $+2$ for
$\CNCBt$ and $-2$ for $\CBCNt$. The latter defects were first
considered theoretically in
Refs.~\onlinecite{Berseneva2011,Berseneva2013}. Our calculated
formation energies are in good overall agreement with those
calculations (apart from the existence of $q=-3$ charge states; see
Ref.~\onlinecite{Bing2011} for a discussion on this point). The
differences with the results of
Refs.~\onlinecite{Berseneva2011,Berseneva2013} can be explained by
significantly larger supercells used in our work, as well as the
electrostatic correction term $E_{\text{corr}}$
[Eq.~\eqref{eq:formation}], which is especially important for higher
charge states. Formation energies of trimers have been first reported
in Ref.~\onlinecite{li2021}, and our calculations are in good
agreement with those results. We will label all defects, the formation
energies of which are shown in Fig.~\ref{fig:C_comp_odd}, as ``odd
carbon clusters''.

\begin{figure}
\includegraphics[width=1\linewidth]{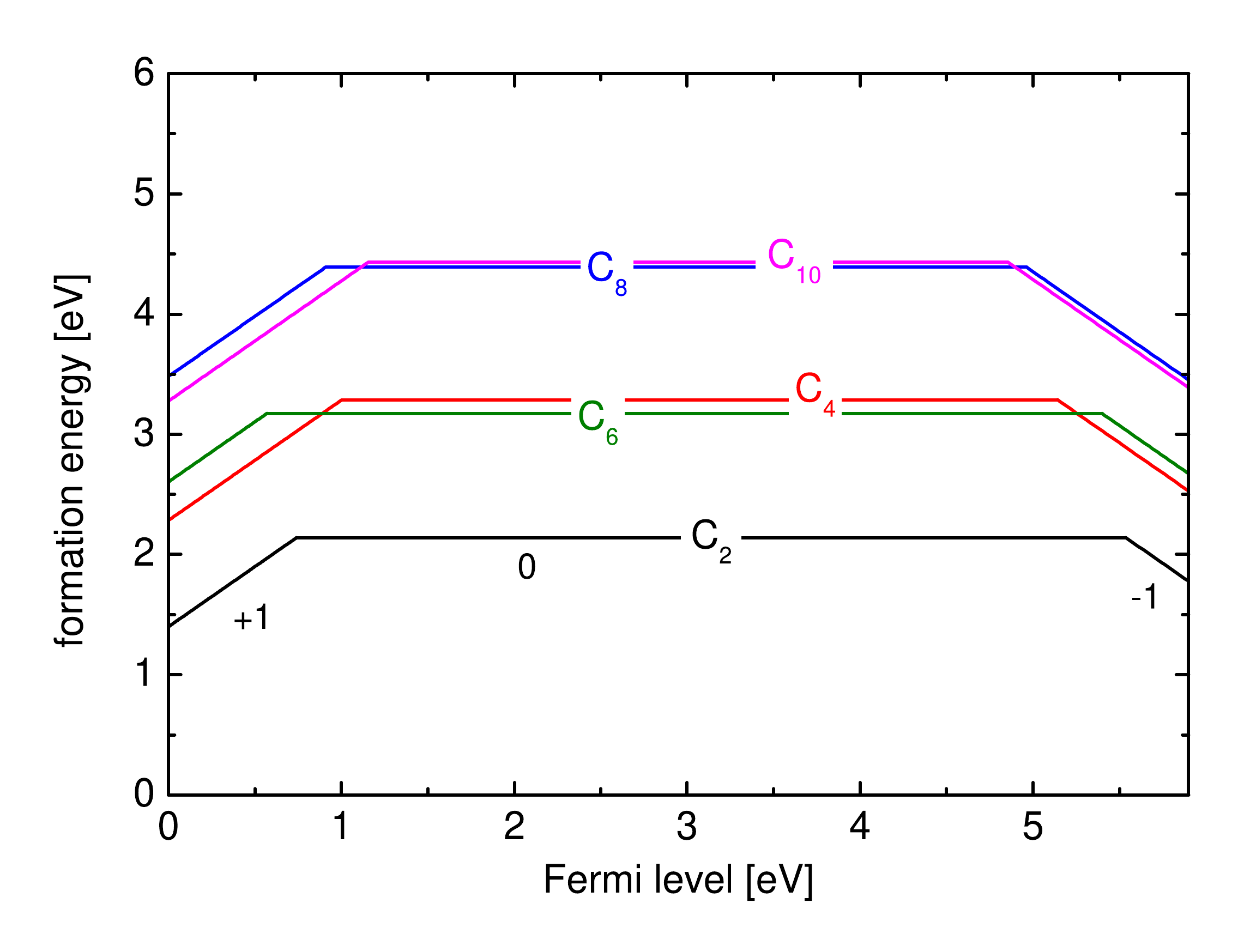}
\caption{Formation energy of even-number carbon clusters as a function
  of the Fermi level. Defects included: carbon dimer $\text{C}_2$,
  tetramer $\text{C}_4$, hexamer $\text{C}_6$, octomer $\text{C}_8$,
  and decamer $\text{C}_{10}$.}
\label{fig:C_comp_even}
\end{figure}

\textit{Even carbon clusters}. Formation energies of carbon clusters
with an equal number of carbon on nitrogen and boron sites are the
same for nitrogen-rich and nitrogen-poor conditions. We will label
these clusters as ``even carbon clusters''. Formation energies for clusters
$\text{C}_2$, $\text{C}_4$, $\text{C}_6$, $\text{C}_8$, and
$\text{C}_{10}$ are shown in Fig.~\ref{fig:C_comp_even}. The result
for the dimer has been previously presented in
Ref.~\onlinecite{mackoit2019}. We find that in the case of all even
carbon clusters, the neutral charge state is the most stable one
throughout most of the bandgap. $+1$ and $-1$ charge states can be
stabilized for Fermi levels close to the band edges. In our
  simulations, we find that even carbon clusters are always
  predominantly charge-neutral for the actual Fermi levels.

Carbon clusters $\text{C}_n$ with $n>3$ can have several non-identical
configurations. Formation energies of the most stable of those are
shown in Figs.~\ref{fig:C_comp_odd}
and~\ref{fig:C_comp_even}. Concentrations of only these most stable
clusters are analyzed in Sec.~\ref{sec:thermodynamics}. For example,
the hexamer $\text{C}_{6}$ can exist either as an open chain or a
closed ring (an analogue of the benzene molecule). The energy of these
two configurations are different: the ``benzene'' configuration has
the lowest energy, while the energy of an open chain is larger. A
slightly different situation is encountered for $\text{C}_4$. It can
exist in either the \emph{cis} or the \emph{trans} configuration, both
shown in Fig.~\ref{fig:defects}. The energies of these two
configurations are nearly equal, and both of them are included in our
analysis. In general, we find that the formation energy of even-number
carbon clusters in the neutral charge state is essentially determined
by the number of C--N and C--B bonds (in the case of BCN
alloys, an equivalent relationship was first pointed out in
Ref.~\onlinecite{mazzoni2006}). This simple chemical picture explains
nearly equal formation energies of both configurations of
$\text{C}_4$. It also explains nearly equal formation energies of
$\text{C}_4$ and $\text{C}_6$ (three pairs of C--N and C--B bonds), as
well as lowest-energy configurations of $\text{C}_8$ and $\text{C}_{10}$ (four pairs), 
evident from Fig.~\ref{fig:C_comp_even}. The $\text{C}_8$ complex with the lowest energy
can be thought of as a carbon ring plus a carbon dimer attached to it.
The lowest-energy configuration of the $\text{C}_{10}$ complex is made
of two carbon rings, an equivalent of the naphthalene molecule. When 
calculating the lattice multiplicity factor $g_D$, we consider only
the lowest-energy configuration or, in the case of tetramers,
the two configurations that have nearly equal energies. Other configurations 
with larger formation energies occur at smaller concentrations in comparison to the most
stable configurations. 

\begin{figure}
\includegraphics[width=1\linewidth]{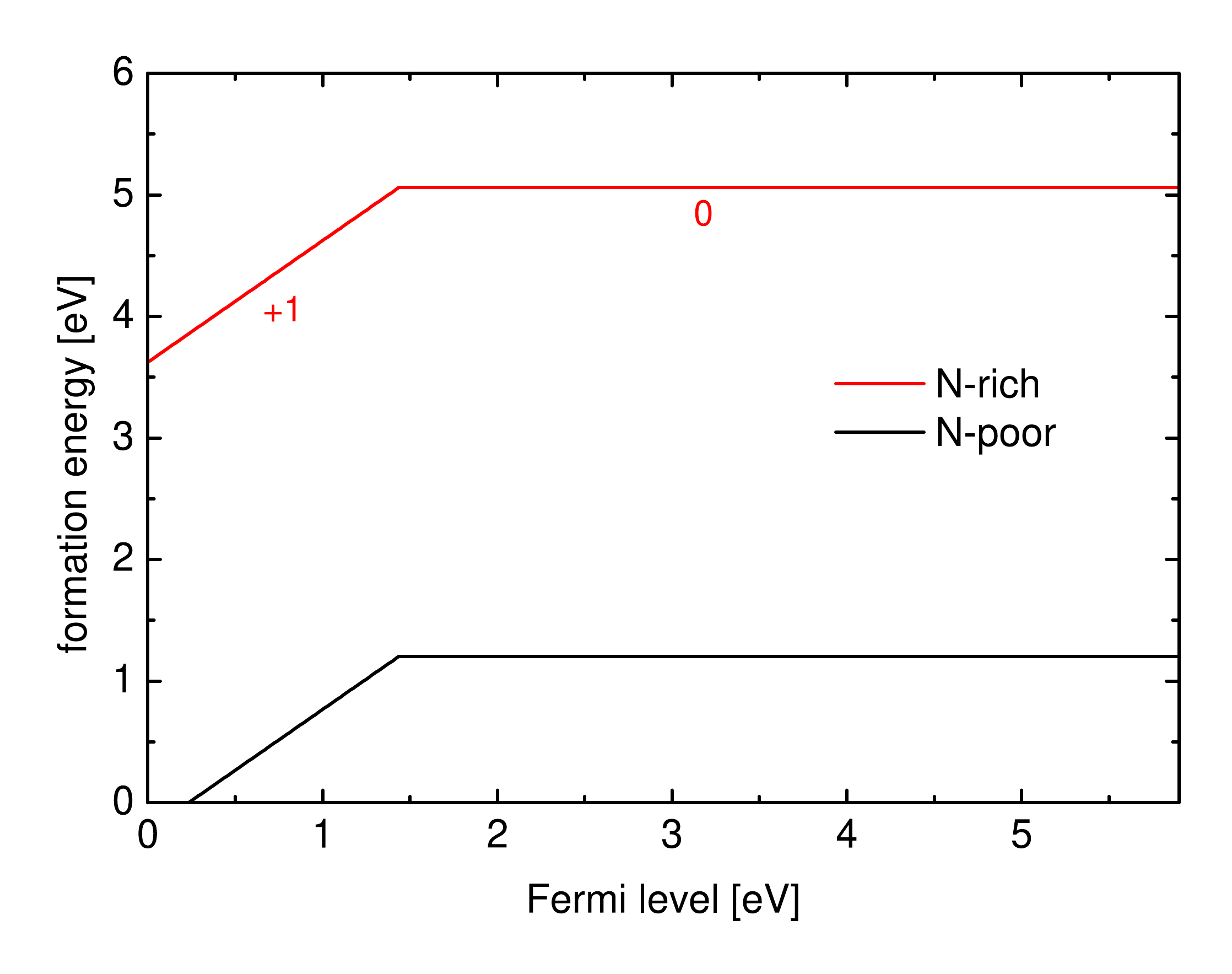}
\caption{Formation energy of the $\CNON$ defect as a function of the
  Fermi level for N-rich and N-poor conditions.\label{fig:formationO}}
\end{figure}

\textit{Oxygen-containing defects.} Formation energies of the $\CNON$
defect are shown in Fig.~\ref{fig:formationO} for both N-rich and
N-poor conditions. Under N-poor conditions, the formation energy of
the neutral complex is very low, just 1.2~eV. Coulomb
attraction between a positively charged oxygen and a negatively
charged carbon explains such a low value. 
Indeed, by comparing the total energy of the separated $\text{C}_\text{N}\mbox{--}\text{O}_\text{N}$
pair and the $\CNON$ defect, we obtain a Coulomb stabilization energy of 1.5 eV.
We can thus anticipate this defect to form in large concentrations whenever both
oxygen and carbon are present under N-poor conditions. The formation
energy is significantly larger under N-rich conditions. Formation
energies of considered oxygen defects not containing carbon are given
in, e.g.  Ref.~\onlinecite{weston2018}.

\textit{Other carbon defects.} Under typical growth temperatures of
hBN, carbon complexes with vacancies and antisites occur at
concentrations smaller than $10^{14}~\mbox{cm}^{-3}$, and their
formation energies are given in the Supplemental
Material~\cite{supp}. The same is true for non-nearest-neighbor $\CB$--$\CN$ pairs, which are also analyzed in
the Supplemental Material~\cite{supp}. As discussed in
Sec.~\ref{sec:thermodynamics}, the only exception is the $\CBVB$
pair. Its formation energy is also given in the Supplemental Material.

\subsection{Equilibrium densities of defects\label{sec:thermodynamics}}

\begin{figure*}
\includegraphics[width=1\textwidth]{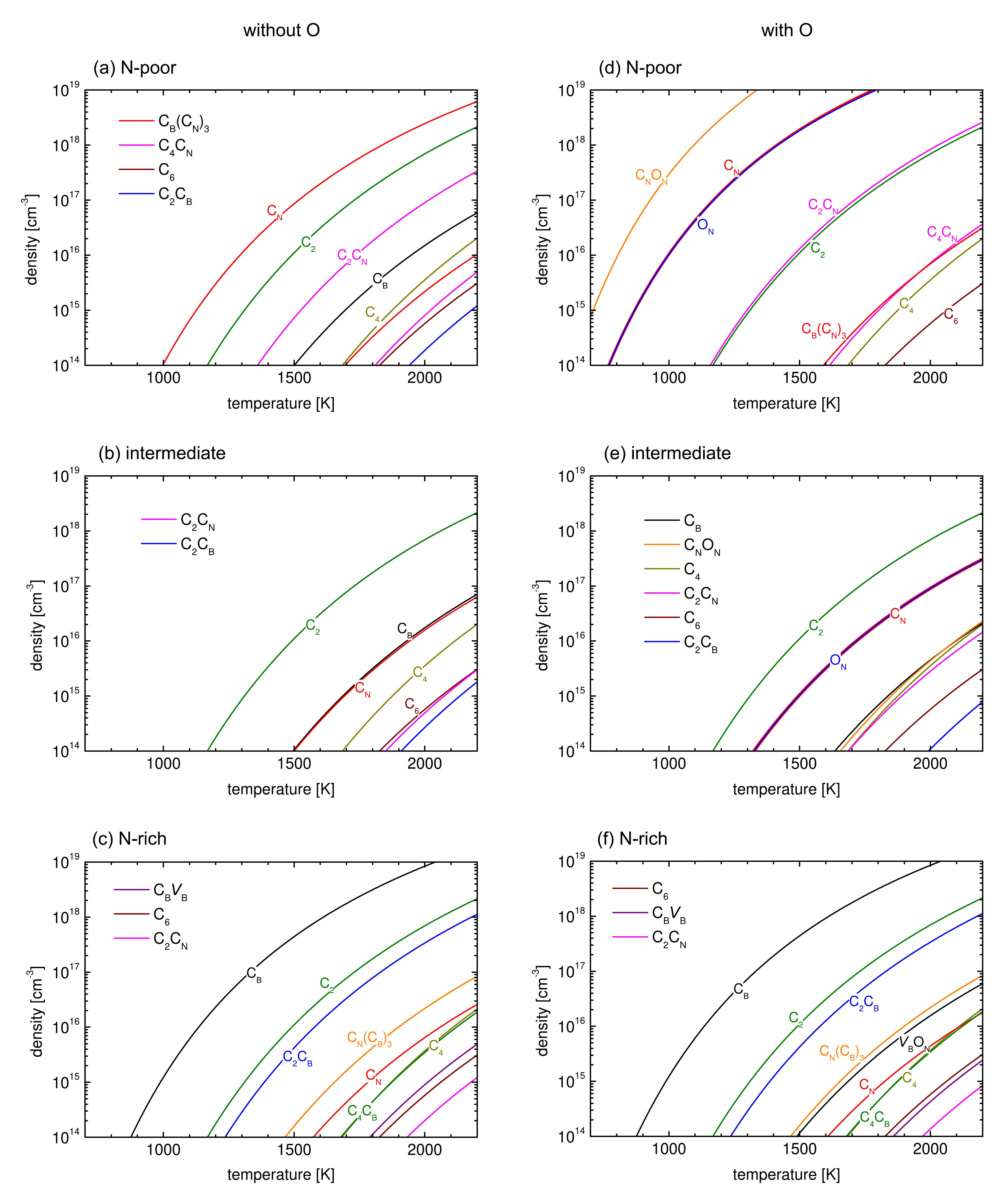}
\caption{Equilibrium concentration of carbon defects as a function of
  temperature without (a--c) and with (d--f) the presence of oxygen during
  growth.  (a) and (d): N-poor conditions; (b) and (e): intermediate
  conditions between N-poor and N-rich; (c) and (f) N-rich conditions.
  Carbon-rich conditions are assumed. These conditions yield the maximum
  density of considered defects. Only defects that attain concentrations 
  $N_D>10^{14}~\mbox{cm}^{-3}$ below temperatures 
  $T=2000$~K are shown.\label{fig:conc1}}
\end{figure*}

\textit{{h}BN with carbon only.} Calculated densities of carbon
defects in the absence of oxygen are shown in
Figs.~\ref{fig:conc1}(a--c) for temperatures up to $T=2200$~K.
This range covers temperatures of the most popular growth techniques
of hBN: 1550--1650~K for molecular beam epitaxy (MBE), about 1600~K
for metalorganic vapor-phase epitaxy (MOVPE), and 1800--2200~K for
bulk growth \cite{Mendelson2020}. Fig.~\ref{fig:conc1}(a) corresponds to N-poor conditions,
(c)~to N-rich conditions, (b)~corresponds to the intermediate
situation exactly at the mid-point between N-rich and
N-poor.  The densities shown in Fig.~\ref{fig:conc1} reflect all charge states
of defects. However, one charge state is often the most
prevalent. We discuss the most prevalent charge state of
defects below, and for illustration purposes, we choose  
the temperature $T=1600$~K. 

Under N-poor conditions, the carbon on the nitrogen site $\CN$ is the
most prevalent defect [Fig.~\ref{fig:conc1}(a)]. 
At $T=1600$~K, the Fermi level is about 1.9~eV above the VBM [Fig.~\ref{fig:C_comp_odd}(a)]. Therefore, $\CN$ is mostly
neutral. The concentration of carbon dimers $\CC$ is about 10 times
smaller. Trimers $\CNCBCN$ (predominantly in the neutral charge state)
also occur in appreciable quantities. These defects are followed by
positively charged $\CB$, neutral triangular defect $\CBCNt$, and neutral clusters
$\text{C}_4$, $\text{C}_6$, and $\CFCN$. Larger carbon clusters and all other
carbon-related defects occur in smaller densities.

The situation under N-rich conditions [Fig.~\ref{fig:conc1}(c)]
approximately mirrors the situation under N-poor conditions with the
substitution $\CN\leftrightarrow\CB$. At $T=1600$~K, the Fermi level
is about 4.7~eV above the VBM, and the defect chemistry is dominated
by neutral $\CB$. Neutral carbon dimers and neutral $\CBCNCB$ occur at
smaller concentrations. These are followed by neutral star-like
defects $\CNCBt$, negatively charged $\CN$, negatively charged
$\CNCBCN$ trimers, and neutral clusters $\text{C}_4$, $\CFCB$, and 
$\text{C}_6$. A noticeable exception to the ``mirror symmetry'' is the
$\CBVB$ defect. This defect occurs in the
$q=-2$ charge state, and its concentration can exceed
$10^{14}~\mbox{cm}^{-3}$ (but only at growth temperatures
$T > 1800$~K).

Under intermediate conditions, the picture is rather different from
the previous two scenarios [Fig.~\ref{fig:conc1}(b)]. At temperature
$T=1600$~K, the Fermi level is 3.3~eV above the VBM, i.e., close to
the mid-gap. We find that carbon dimers $\CC$ become the most dominant
defects (their formation energies are independent of chemical
potentials of boron and nitrogen). The concentration of single
substitutionals is about two orders of magnitude smaller (though still quite
appreciable). Single substitutionals are followed by $\text{C}_4$ and
$\text{C}_6$. Like in the case of dimers, the formation energies of
$\text{C}_4$ and $\text{C}_6$ clusters are independent of boron and
nitrogen chemical potentials (within the constraint of equilibrium
growth, $\mu_B+\mu_C=\mu_{BN}$). Formation energies of these two
defects are similar (Fig.~\ref{fig:C_comp_even}), and larger densities
of tetramers are explained by the difference in lattice multiplicity
factors $g_D$ (6 for tetramers vs.~$1/2$ for hexamers).  Under these
intermediate conditions, carbon trimers appear in much smaller
concentrations than under ``extreme'' N-poor or N-rich conditions. The
concentration of triangular star-like defects falls significantly
below our chosen threshold of $10^{14}~\mbox{cm}^{-3}$.

\textit{{h}BN with carbon and oxygen.} The presence of oxygen changes
the overall picture. The concentration of carbon- and
oxygen-containing point defects is shown in
\mbox{Fig.~\ref{fig:conc1}(d--f)}.  The effect of oxygen is most
pronounced under N-poor conditions [Fig.~\ref{fig:conc1}(d)]. It
manifests itself (i)~via the large concentration of oxygen-containing
defects and (ii)~via a change in concentrations of carbon defects not
containing oxygen. Due to the very low formation energy of the $\CNON$
defect under N-poor conditions (Fig.~\ref{fig:formationO}), $\CNON$
occurs in concentrations exceeding $10^{19}~\mbox{cm}^{-3}$ for growth
temperatures $T>1300$~K. This defect occurs primarily in the neutral
charge state and does not influence the position of the Fermi
level. However, the Fermi level is affected by the incorporation of
oxygen donors $\ON$ that also occur in substantial concentrations. In
comparison to the system without oxygen, the Fermi level is pushed up
by more than 1~eV [Fig.~\ref{fig:C_comp_odd}(a)]; $E_F$ is now
determined by the equilibrium between positively charged donors $\ON$
and negatively charged acceptors $\CN$. As a result, the concentration
of $\CN$ also increases (due to an increase of negatively charged
species; the concentration of neutral $\CN$ is the same as in the case
without oxygen). The raising of the Fermi level does not affect the
concentration of even carbon clusters. However, higher Fermi levels
increase the concentration of negatively charged odd carbon clusters
with surplus $\CN$ units and decrease the concentration of positively
charged carbon clusters with surplus $\CB$ units. Most importantly, we
find that the concentration of carbon trimers $\CNCBCN$ becomes quite
appreciable, very close to that of dimers. Similarly, concentrations
of star-like defects $\CBCNt$ and pentamers
C$_4 \text{C}_{\text{N}}$ are boosted.\looseness=1

In contrast to N-poor conditions, oxygen does not significantly affect
the formation of carbon defects under N-rich conditions
[Fig.~\ref{fig:conc1}(f)]. The densities of carbon defects remain
similar to the case without oxygen [Fig.~\ref{fig:conc1}(c)].

Under intermediate conditions between N-rich and N-poor
[Fig.~\ref{fig:conc1}(e)], the situation resembles that with no oxygen. The presence of oxygen raises the Fermi level by about
0.3~eV, and $E_F$ nearly reaches $\CB$ defect's $(+/0)$ charge-state
transition level. The consequence of this is a smaller concentration
of $\CB$ defects, and charge neutrality is established by the
equilibrium between positively charged $\ON$ donors and negatively
charged carbon acceptors $\CN$. A modest increase of the Fermi level
slightly increases the concentration of carbon defects with surplus
$\CN$ units and decreases the concentrations of those with surplus
$\CB$ units. Regarding carbon defects, one could conclude that the
influence of oxygen is not particularly strong in this regime
[cf.~Fig.~\ref{fig:conc1}(b) and (e)].

%%%%%%%%%%%%%%%%%%%%%%%%%%%%%%%%%%%%%%%%%%%%%%%%%%%%%%%%%%%%%%%%%%%%%%
%%%%%%%%%%%%%%%%%%%%%%%%%% EXPERIMENT %%%%%%%%%%%%%%%%%%%%%%%%%%%%%%%%
%%%%%%%%%%%%%%%%%%%%%%%%%%%%%%%%%%%%%%%%%%%%%%%%%%%%%%%%%%%%%%%%%%%%%%

\section{Defect thermodynamics and single-photon emission in
  carbon-doped hBN\label{sec:expt}}

In this section, we discuss the findings of
Ref.~\onlinecite{Mendelson2020} in light of the results presented in
Sec.~\ref{sec:thermodynamics}. We will mostly focus on MOVPE samples
that are grown at ${\sim}1600~\text{K}$ \cite{Mendelson2020}. For
these samples, carbon was introduced via the triethylborane (TEB)
precursor, the flow rate of which could be varied. The resulting epitaxial layers
were about $h=40$~nm thick. Altering the flow rate of the carbon-carrying
precursor effectively changes the chemical potential of carbon. The highest
chemical potential that can be achieved is determined by the onset of
formation of bulk inclusions of graphite (see
Sec.~\ref{sec:validity}). This is the carbon-rich limit, whereby the
chemical potential of carbon is determined by the energy of one C atom
in graphite. It was exactly this limit that was assumed in the
calculation of formation energies and concentrations in
Sec.~\ref{sec:formation}. This limit determines the maximum possible
concentration of carbon-related point defects. When the flow of the
precursor is lowered, carbon chemical potential decreases, which
increases the formation energy of carbon-related defects, leading to
their smaller concentrations.

In Ref.~\onlinecite{Mendelson2020}, the authors could detect isolated
carbon-related SPEs, emitting in the energy range
$1.6\mbox{--}2.1$~eV, for the lowest flow rate of the precursor TEB
($10~\mu \text{mol}/\text{min}$). However, single emitters could no
longer be identified for higher flow rates of the TEB precursor
($\ge 20~\mu \text{mol}/\text{min}$), as the density of
emitters became too large.  The emitters can no longer be identified
as single ones when their lateral density becomes larger than one
emitter per diffraction spot of the laser. 
The diameter of the laser spot can be assumed to be equal to the diameter of the Airy disk,
$d=1.22 \lambda / \rm{NA}$, where $\lambda$ is the wavelength of the laser
and $\rm{NA}$ is numerical aperture. Using the parameters of Ref.~\cite{Mendelson2020},
$\lambda=532$ nm and $\rm{NA}=0.95$, we obtain $d\approx 680$ nm. 
The total volume per one emitter is then smaller than $\pi d^2 h /4$,
yielding 3D concentrations larger than $0.7 \times 10^{14}~\mbox{cm}^{-3}$. 

This means that emitters observed
in Ref.~\onlinecite{Mendelson2020} are defects that \textit{can} occur
in concentrations larger than about $N_{D} =
10^{14}~\mbox{cm}^{-3}$. This was exactly the number cited in
Sec.~\ref{sec:formation} and chosen as our criterion. However, one
must keep in mind that it is not entirely clear whether the maximum
flow rate of the carbon precursor used in
Ref.~\onlinecite{Mendelson2020} indeed corresponds to the carbon-rich
limit. If it does not, then the carbon-rich limit would yield even
higher defect concentrations than the ones obtained in
Ref.~\onlinecite{Mendelson2020}. This makes our chosen criterion
$N_{D} = 10^{14}~\mbox{cm}^{-3}$ a good measure regarding possible
defect centers to explain the results of
Ref.~\onlinecite{Mendelson2020}.

\subsection{hBN with carbon only}

Let us now analyze the results of our calculations shown
in Fig.~\ref{fig:conc1} and compare them with the results of Ref.~\onlinecite{Mendelson2020}.
We will first consider the situation without oxygen present.

\emph{N-poor conditions.} In the case of N-poor conditions
[Fig.~\ref{fig:conc1}(a)], we see that only a few carbon-related
defects can occur at concentrations larger than
$10^{14}~\mbox{cm}^{-3}$ for growth temperatures $T=1600$~K. These
defects are: $\CN$, the dimer $\CC$, the trimer $\CNCBCN$ (all
charge-neutral), and positively charged $\CB$. The neutral triangular
defect $\CBCNt$ also approaches threshold
concentrations. Historically, $\CN$ was
associated~\cite{Katzir1975} with the ubiquitous 4.1~eV luminescence
line in hBN. However, density
functional theory calculations showed that this attribution was not
correct~\cite{weston2018}. In recent theoretical work, Auburger and
Gali calculated the ZPL energy of the neutral $\CN$ to be about
2.5~eV~\cite{Auburger2021}, not far from the experimentally measured
values of visible SPEs. The optical transition occurs from a localized defect level to a
perturbed bulk state, stabilized by electron-hole interaction energy of $\approx 0.8$ eV. 
This defect, however, possesses a $D_{3h}$ point group symmetry. As a result, there should not be a fixed
emission dipole in the hBN plane for dipole-allowed optical
transitions. Instead, the emission could only have random polarization 
in the hBN plane. This is in contradiction with experiments that show a
fixed dipole~\cite{Exarhos2017}. However, local electric or strain
fields can lower the symmetry and lock the direction of the transition
dipole moment with respect to the underlying lattice.

There is firm theoretical evidence that the carbon dimer defect emits in the UV region, 
likely being the cause of the 4.1~eV emission~\cite{mackoit2019,winter2021,muechler2021}. Our
thermodynamic modelling indicates that the concentration of dimers
should exceed $10^{16}~\text{cm}^{-3}$ in MOVPE samples,
meaning that dimers are indeed ubiquitous. Regarding the optical signature of the neutral trimer
$\CNCBCN$, it was recently proposed that its theoretical luminescence
lineshape strongly resembles that of some 1.6--2.1~eV
SPEs~\cite{Jara2021}. Calculations of Refs.~\onlinecite{Jara2021,Auburger2021} yield
the value of 1.62~eV for the ZPL energy, indeed
close to the experimental range. Our
thermodynamical calculations confirm that this defect is a contender
to explain single-photon emission in the IR/visible energy range.

Positively charged $\CB$ defects that occur in smaller concentrations
than neutral trimers $\CNCBCN$ do not have filled defect states and
therefore should not have an internal defect transition.

Finally, to the best of our knowledge, optical properties of the
triangular defect $\CBCNt$ have not been investigated so far. Whenever
this defect occurs in significant concentrations (``extreme'' N-poor
conditions), it is the neutral charge state that is the most
prevalent. The ground state of the defect has spin $S=1$.
A preliminary analysis of the electronic structure of this defect
reveals that the excited state is of multi-determinant nature. As a result,
it is not straightforward to obtain the ZPL energy using so-called
delta-self-consistent-field ($\Delta$SCF) calculations \cite{Jones1989} that
are frequently used for defects. However, the nature of the excited state is expected to be similar to that of the $\CN$
defect, analyzed in Ref.~\onlinecite{Auburger2021}. Thus, we can make an estimate of the ZPL energy of this
defect by determining the energy needed to excite a hole to a valence band
and taking into consideration the electron-hole interaction energy of $\approx 0.8$ eV.
In this way we obtain a ZPL energy of $\approx 2.2$ eV, very close
to the energy of visible emitters in hBN. Due to its triplet ground state, 
the $\CBCNt$ defect is potentially an interesting center in hBN,
worth further investigation. From the thermodynamical standpoint, our
results show that this defect can occur in sufficient concentrations
under N-poor conditions. Based both on the thermodynamical modelling
and an estimate of the ZPL energy, we cannot exclude this defect as
contributing to single-photon emission in the 1.6--2.1~eV energy
range. However, like the $\CN$ defect, the $\CBCNt$ center possesses
$D_{3h}$ symmetry. Thus, optical polarization of emission could have a random
orientation in the hBN plane, but should not have a fixed in-plane 
transition dipole moment in the absence of other perturbations.

\emph{N-rich conditions.} The situation for N-rich conditions
[Fig.~\ref{fig:conc1}(c)] is in a way symmetrical to that of N-poor
conditions under substitution $\CN \leftrightarrow \CB$. The most
dominant defect is neutral $\CB$.  Auburger and Gali obtained the ZPL
energy of about 1.7 eV for this defect, falling in the experimental range \cite{Auburger2021}. Luminescence corresponds to a transition of the electron from a perturbed conduction band state, stabilized by electron-hole interaction, to a localized defect state. This defect is followed by the carbon dimer and the trimer $\CBCNCB$. The calculated
optical signature of this trimer is similar to that of
$\CNCBCN$~\cite{Jara2021}. The calculations of
Ref.~\cite{Jara2021} yield the ZPL value of 1.65~eV,
while a smaller value of 1.36~eV has been obtained in Ref.~\cite{Auburger2021}. 
These results hint that the $\CBCNCB$ defect is a less likely contender to explain
emission in the 1.6--2.1~eV energy range. However, keeping in mind
possible errors in the evaluation of optical transition energies,
based on thermodynamics alone, we expect that the $\CBCNCB$ trimer can
also be a potential candidate to explain single-photon emission in the
red and near-infrared.

The triangular defect $\CNCBt$ occurs in rather large concentrations
($10^{15}~\text{cm}^{-3}$) at growth temperatures of 1600~K. Like in
the case of the $\CBCNt$ defect, we find that the neutral charge state
is the most relevant one. The ground state of $\CNCBt$ is
spin-triplet ($S=1$). We can estimate the ZPL energy of this defect similarly to
the $\CBCNt$ defect. We obtain the energy of $\approx 2.0$ eV. Due to its triplet ground state, the $\CNCBt$ 
triangular defect is also a potentially interesting system for quantum applications. Based on
thermodynamical modelling and the estimate of the ZPL energy, we 
suggest this defect as a potential candidate to explain single-photon
emission in the 1.6--2.1~eV energy range. These preliminary estimates
call for a more rigorous investigation of the electronic structure and
optical transitions of both triangular star-like defects, which should
be the focus of future work.

\emph{Intermediate conditions.} At conditions intermediate between
N-rich and N-poor, the concentration of trimers and triangular defects
falls significantly below $10^{14}~\text{cm}^{-3}$ and carbon
chemistry is dominated by carbon dimers. Apart from dimers and single
substitutional defects, no other carbon-related defects form in large
enough concentrations.

\emph{Bulk samples.} Bulk samples are grown at temperatures
$T=1800\mbox{--}2200$~K~\cite{Taniguchi2007}. Figure~\ref{fig:conc1}
indicates that under these conditions, the formation of larger carbon
clusters, in particular, tetramers $\text{C}_4$ 
and hexamers $\text{C}_6$, becomes more pronounced and exceeds
$10^{14}~\text{cm}^{-3}$. In MOVPE samples grown at lower temperatures
the concentration of these larger clusters should be about $10^{13}~\text{cm}^{-3}$.
In addition to these defects, carbon pentamers $\CFCN$ are quite common
under N-poor conditions, while $\CFCB$ show up under N-rich conditions.
The existence of carbon hexamers, along with dimers, was indeed confirmed 
in Ref.~\onlinecite{Krivanek2010}, where authors used annular dark-field electron microscopy to image hBN
monolayers exfoliated from bulk samples. 

Optical emission energies of these defects are not known. In the case of the $\text{C}_4$ defect,
we determine the ZPL energy using the aforementioned $\Delta$SCF calculations.
Since the excited state is an spin-singlet, we have to apply a singlet
correction as has been discussed in, e.g., Ref.~\cite{mackoit2019}.
We obtain ZPL energy of 3.10 eV for the \emph{trans} isomer and
3.45 eV for the \emph{cis} isomer. Unfortunately $\Delta$SCF calculations 
cannot be straightforwardly applied for the $\text{C}_6$ defect due its more 
complicated electronic structure. A simple estimate can be made based on our calculated 
Kohn-Sham energy levels of the $\text{C}_6$ defect and the comparison with the dimer defect, 
for which ZPL energies can be calculated. Using this approximation, we can estimate that ZPL
energy of the carbon hexamer should be about 4.0~eV. Similarly, we estimate that the ZPL energy of the 
$\CFCN$ pentamer should be about 0.15 eV smaller than the corresponding energy of the $\CNCBCN$ trimer,
and that of the $\CFCB$ pentamer by about 0.15 eV smaller than the ZPL energy of the $\CBCNCB$ trimer.

In addition to these
larger carbon clusters, doubly negatively charged $\CBVB$ complex can
appear in sufficient concentrations in bulk samples grown under N-rich
conditions. In their
paper, Korona and Chojecki \cite{Korona2019} studied optical
transitions of the neutral center and suggested ZPL energies of about
1~eV. However, we find that the doubly negatively charged center,
which is the relevant charge state, does not have an internal optical
transition, as all of the defect states in the bandgap are
filled. Therefore, this defect is an unlikely candidate to contribute
to single-photon emission in the energy range 1.6--2.1~eV.

\subsection{hBN with carbon and oxygen}

As discussed in Sec.~\ref{sec:thermodynamics}, the presence of oxygen
most strongly affects carbon defects concentration under N-poor
conditions. The largest difference is the occurrence of the $\CNON$
defect, which can form in amounts exceeding any other carbon and
oxygen defects. The emission energy of
this defect is not firmly established. It was
proposed~\cite{vokhmintsev2019} that this defect causes the
afore-mentioned 4.1~eV luminescence line, rivalling the carbon dimer
hypothesis~\cite{mackoit2019}. The concentrations of these defects,
however, are drastically different in different hBN samples.  $\CNON$
is clearly dominant under ``extreme'' N-poor conditions, but the
density of this defect falls quickly with the increase of the nitrogen
chemical potential. At variance, the concentration of carbon dimers
depends only on the chemical potential of carbon and is the same under
N-poor and N-rich conditions. This distinction is an important handle
that could potentially help differentiate between the two defects.

The presence of oxygen does not affect our conclusions regarding SPEs
in the energy range 1.6--2.1~eV. As already discussed, oxygen causes
increased densities of $\text{C}_{\text{N}}$ defects and odd carbon
clusters with surplus $\text{C}_{\text{N}}$ units (under N-poor
conditions). However, no other defects apart from the ones analyzed
above occur at concentrations higher than $10^{14}~\text{cm}^{-3}$.
\vspace{-0.5em}

\section{Validity of the formation energy
  formalism\label{sec:validity}}
\vspace{-0.5em}

Our analysis above relied on the use of the formation energy
formalism, outlined in Sec.~\ref{sec:th}. The key aspect of formalism
is that an external carbon source sets the carbon chemical potential
$\mu_{\text{C}}$. Setting $\mu_{\text{C}}$ to the energy of carbon in
graphite defines the maximum value of the carbon chemical
potential. These are carbon-rich conditions that yield the maximum
possible concentration of carbon-containing defects. Assuming a fixed
chemical potential seems a natural choice when modeling epitaxial
growth, e.g., MOVPE or MBE.

At variance, bulk growth occurs in closed crucibles with a fixed
amount of different chemical species, in our case, boron, nitrogen,
carbon (and oxygen or other impurities, depending on conditions). In
this case, the formation energy formalism to compute concentrations of
point defects can still be applied. However, chemical potentials,
including that of carbon, are not determined by external sources but
are governed by the total amount of different chemical species in the
growth chamber.

Let us consider hBN with a certain density of carbon $N_{\text{C}}$ at
a temperature $T$. When $N_{\text{C}}$ is sufficiently low, all the
carbon (under thermodynamic equilibrium conditions) is in the form of
point defects. The maximum amount
of carbon that can exist in the form of point defects is given by the
expression:
\begin{equation}
N_{\text{C},\text{max}}(T)=\sum_{D} n_{\text{C}} N_{D,\text{max}}(T).
\label{eq:solubility}
\end{equation}
Here, the sum runs over all carbon-containing
defects. $N_{D,\text{max}}$ is the equilibrium density of defect $D$
in all charge states calculated for carbon-rich conditions,
$N_{D,\text{max}}(T)=\sum_{q}N_{D,q,\text{max}}(T)$ (the same as given
in Fig.~\ref{fig:conc1}; here we added the subscript ``max'' to
emphasize that these are carbon-rich conditions).
$N_{\text{C},\text{max}}(T)$ can thus be thought of as the solubility 
limit of carbon in hBN when carbon is in the form of point defects.

When the actual amount of carbon is below the solubility limit, the
procedure to determine the value of the carbon chemical potential
$\mu_{\text{C}}$ is as follows. Formation energies of point defects
are determined by Eq.~\eqref{eq:formation}. The concentrations of
defects are given by Eq.~\eqref{eq:ND}. Carbon chemical potential is
determined by imposing the condition
$\sum_{D} n_{\text{C}} N_{D}(T)=N_{\text{C}}$. The three equations
have to be solved self-consistently (in addition to the equation to
determine the electron chemical potential), which in the end yields
both the actual value of $\mu_{\text{C}}$ and the concentration of
different carbon species. The procedure has been previously applied to other systems \cite{Windl2021}
and is illustrated quantitatively in the Supplemental Material~\cite{supp}. When the amount of carbon
exceeds $N_{\text{C},\text{max}}(T)$, the carbon chemical potential
corresponds to the carbon-rich limit, and the concentrations of point
defects are the ones that are shown in Fig.~\ref{fig:conc1}.
The surplus carbon condenses into graphene/graphite islands.

\begin{figure}
  \centering
\includegraphics[width=1\linewidth]{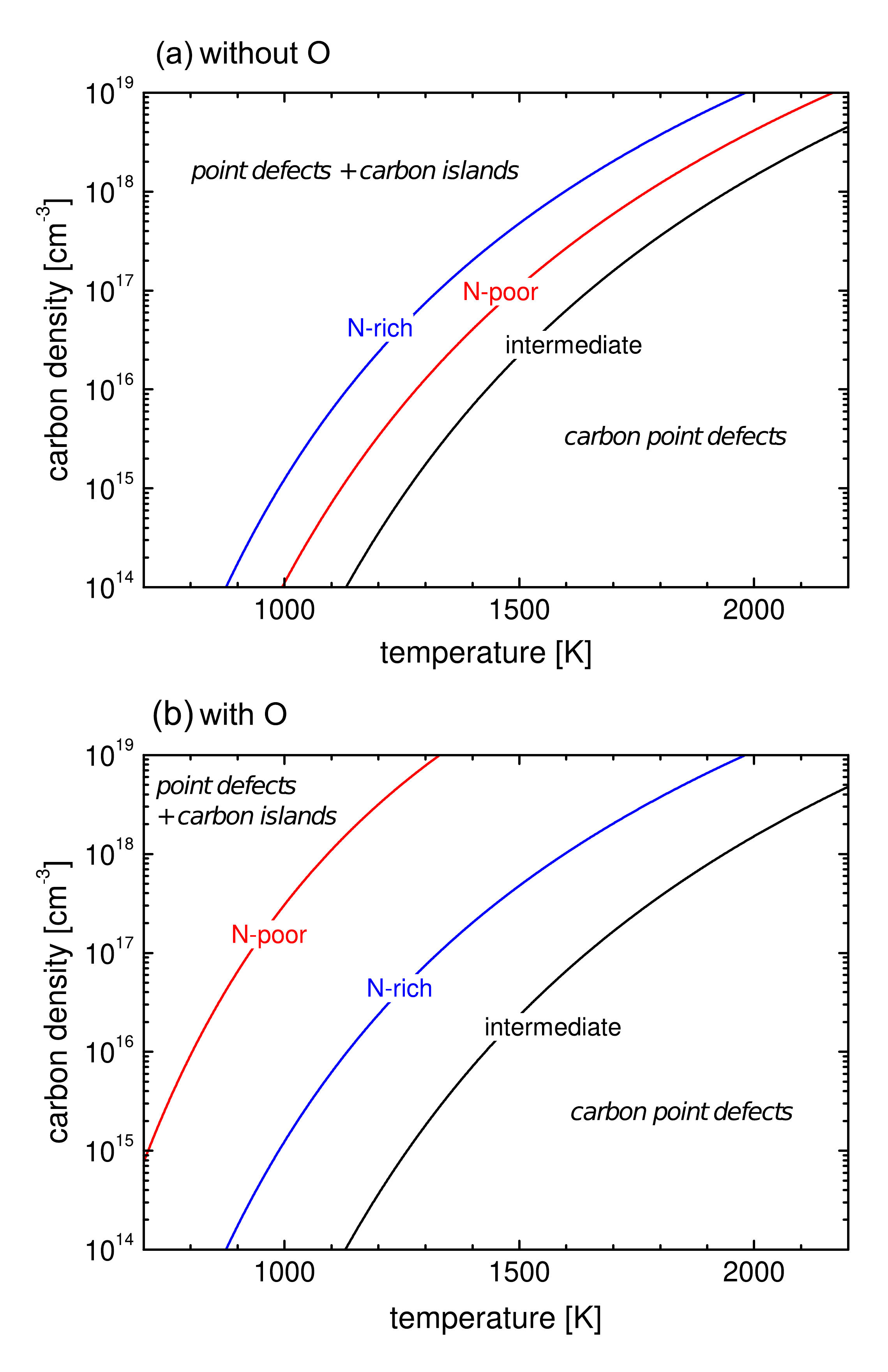}
\caption{Solubility limit of carbon in hBN as a function of
  temperature for N-poor conditions, intermediate conditions between
  N-poor and N-rich, and N-rich conditions.
  The solubility limit corresponds to the maximum density of carbon
  atoms that can be incorporated into hBN in the form of point defects. 
  For carbon densities above the solubility limit, carbon starts
  to condense into graphite islands. 
  \label{fig:sol}}
\vspace{-0.5em}
\end{figure}

The calculated solubility limits of carbon in hBN for different
conditions, calculated using Eq.~\eqref{eq:solubility}, are shown in
Fig.~\ref{fig:sol}. Fig.~\ref{fig:sol}(a) depicts the solubility limit
without the presence of oxygen. The solubility limit is slightly
larger for ``extreme'' N-rich and N-poor conditions and smaller for
intermediate conditions. The situation is very different when oxygen
is present, shown in Fig.\ref{fig:sol}(b). The solubility limit is not
much affected by oxygen under N-rich and intermediate conditions but,
in comparison to oxygen-free growth, is significantly increased for
N-poor conditions. This is mostly because of a very large
concentration of $\CNON$ defects.

Fig.~\ref{fig:sol} can also be viewed as the phase diagram of carbon
in hBN in the dilute limit. In this point of view, the solubility
limit is the binodal curve, above which the condensation of carbon
ensues.

%%%%%%%%%%%%%%%%%%%%%%%%%%%%%%%%%%%%%%%%%%%%%%%%%%%%%%%%%%%%%%%%%%%%%%
%%%%%%%%%%%%%%%%%%%%%%% CONCLUSIONS %%%%%%%%%%%%%%%%%%%%%%%%%%%%%%%%%%
%%%%%%%%%%%%%%%%%%%%%%%%%%%%%%%%%%%%%%%%%%%%%%%%%%%%%%%%%%%%%%%%%%%%%%

\begin{table*}
  \caption{Carbon-related defects with maximum equilibrium
    densities. Densities (in $\text{cm}^{-3}$) correspond to growth temperatures
    $T=1600$ K and no oxygen present during growth. ZPL energies $E_{\text{ZPL}}$ of internal
    defect transitions are in eV. Values of $E_{\text{ZPL}}$ are either
    taken from published theoretical calculations (approximated to within 0.1 eV), 
    calculated in this paper, or estimated as described in the text (the latter case
    is distinguished by a symbol ``$\approx$'' before a number). %The error bars for the ZPL
    %values of some defects can be as high as
    %0.5~eV~\cite{muechler2021}. 
    Defects are charge-neutral unless explicitly stated.}

 \begin{ruledtabular}
   \begin{tabular}{c c c c c c c c c c}
     \multicolumn{5}{c}{N-poor conditions } & \multicolumn{5}{c}{N-rich conditions }\\
     \cline{1-5} \cline{6-10}
     \\[-2ex]    
     defect & max.~density & symmetry & spin & $E_{\text{ZPL}}$ & defect & max.~density & symmetry & spin & $E_{\text{ZPL}}$ \\
     \cline{1-5} \cline{6-10}    
     \\[-1.5ex]
  
   $\text{C}_{\text{N}}$                      & $2 \times 10^{17}$ & $D_{3h}$ & 1/2 & 2.5 \cite{Auburger2021}                         & $\text{C}_{\text{B}}$                      & $1 \times 10^{18}$ & $D_{3h}$ & 1/2 & 1.7 \cite{Auburger2021}                          \\
   $\text{C}_2$                               & $3 \times 10^{16}$ & $C_{2v}$ & 0   & 4.3 \cite{mackoit2019}, 4.1 \cite{Auburger2021} & $\text{C}_2$                               & $3 \times 10^{16}$ & $C_{2v}$ & 0   & 4.3 \cite{mackoit2019}, 4.1 \cite{Auburger2021} \\
   $\text{C}_2\text{C}_{\text{N}}$            & $2 \times 10^{15}$ & $C_{2v}$ & 1/2 & 1.6 \cite{Jara2021,Auburger2021}                         & $\text{C}_2\text{C}_{\text{B}}$            & $1 \times 10^{16}$ & $C_{2v}$ & 1/2 & 1.7 \cite{Jara2021}, 1.4 \cite{Auburger2021}                         \\
   $\text{C}_{\text{B}}^+$                    & $3 \times 10^{14}$ & $D_{3h}$ & 0   & $-$                                              & $\text{C}_{\text{N}}(\text{C}_{\text{B}})_3$ & $5 \times 10^{14}$ & $D_{3h}$ & 1   & $\approx 2.0$                               \\
   $\text{C}_\text{B}(\text{C}_{\text{N}})_3$ & $3 \times 10^{13}$ & $D_{3h}$ & 1   & $\approx 2.2$                               & $\text{C}_{\text{N}}^-$                    & $1 \times 10^{14}$ & $D_{3h}$ & 0   & $-$                                              \\
   $\text{C}_4$                               & $3 \times 10^{13}$ & $C_s$    & 0   & 3.10 (\emph{trans}), 3.45  (\emph{cis})                                   & $\text{C}_4$                               & $3 \times 10^{13}$ & $C_s$    & 0   & 3.10 (\emph{trans}), 3.45  (\emph{cis})                                      \\
 %  $\text{C}_6$                               & $6 \times 10^{12}$ & $D_{3h}$    & 0   & $\approx 4.0$ & $\text{C}_6$ & $6 \times 10^{12}$ & $D_{3h}$    & 0   &    $\approx 4.0$  \\

 \end{tabular}
 \label{Table1}
 \end{ruledtabular}
\end{table*}

\section{Conclusions\label{sec:conc}}

In this work, we presented the thermodynamic analysis of carbon defects
in hexagonal boron nitride. Our particular focus was the relevance of
these results to single-photon emission in hBN, where the involvement
of carbon defects was reported~\cite{Mendelson2020}.

Regarding carbon defects themselves, the conclusions of our work are
the following:

1. Under most growth conditions, when no oxygen is present, single
substitutional defects $\CN$ and $\CB$, as well as carbon dimers \CC,
are carbon defects that occur in the largest
concentrations. Theoretical calculations suggest that the ZPL energy
of the $\CC$ defect is about 4.1--4.3~eV~\cite{mackoit2019,Auburger2021}, while ZPL
energies of $\CB$ and $\CN$ defects are around 1.7~eV and 2.5 eV,
respectively~\cite{Auburger2021}. From the perspective of thermodynamics,
the latter two defects are potential candidates
to explain single-photon emission in the energy region 1.6--2.1~eV,
observed in Ref.~\cite{Mendelson2020}.

2. Carbon trimers $\CBCNCB$ and $\CNCBCN$ occur in smaller
concentrations than single substitutionals and dimers. However, their
densities can be larger than $10^{14}~\mbox{cm}^{-3}$ under certain
growth conditions. In particular, $\CBCNCB$ defects should be abundant
under N-rich conditions, while $\CNCBCN$ should be commonplace under
N-poor conditions. Recent theoretical calculations showed that the ZPL
energies of these defects are smaller than
1.7~eV~\cite{Jara2021,Auburger2021}.  Based on thermodynamical
analysis and considering possible errors in evaluating optical
transition energies, we infer that trimers can be deemed likely
candidates to explain single-photon emission in the red and
near-infrared spectral regions. Notwithstanding, the concentration of
trimers is expected to be significantly smaller under intermediate
conditions between N-rich and N-poor. Changing the chemical potentials
of boron and nitrogen is thus a handle that can strongly affect the
incorporation of these defects.

3. Triangular star-like defects
$\text{C}_{\text{B}} (\text{C}_{\text{N}})_3$ and
$\text{C}_{\text{N}} (\text{C}_{\text{B}})_3$ can also occur in
concentrations approaching $10^{14}~\mbox{cm}^{-3}$, but only under
appropriate ``purely'' N-rich or N-poor conditions. For small
deviations from these limit conditions, the formation energies of
these defects increase substantially, and their concentration
plummets. Neutral charge states of these defects are the relevant
ones.  $\text{C}_{\text{B}} (\text{C}_{\text{N}})_3$ is expected to
have the ZPL energy of $\approx 2.2$~eV, while
that of the $\text{C}_{\text{N}} (\text{C}_{\text{B}})_3$ defect should be $\approx 2.0$~eV. 
In contrast to other carbon defects discussed in this paper, both 
star-like defects have a triplet ground state ($S=1$)~\cite{Berseneva2011}, 
which makes them potentially useful for applications such as quantum sensing. 
Star-like defects are therefore
interesting systems that require further studies.
\goodbreak

4. Carbon tetramers $\text{C}_4$ and hexamers $\text{C}_6$ are also
rather prominent defects. Their concentration should be around
$10^{13}~\mbox{cm}^{-3}$ in MBE and MOVPE samples grown under
carbon-rich conditions but can surpass $10^{14}~\mbox{cm}^{-3}$ in
bulk hBN grown at about 2000~K. Carbon hexamers are expected
to have ZPL energies of $\approx 4.0$~eV, i.e., very similar to carbon dimers.
Computed ZPL energies of the carbon tetramer is $3.10-3.45$~eV (depending on its configuration). 
The fact that such larger atomic clusters have sufficiently low formation energies and can 
form in significant amounts is a rather special situation for point defects. The existence 
of these clusters, as demonstrated experimentally for $\text{C}_6$ \cite{Krivanek2010}, 
should be attributed to an intermediate position of carbon between
boron and nitrogen in the periodic table, as well as to similar crystal
structure of graphite and hBN.

5. Among the defects that are neither single carbon substitutionals
nor carbon clusters, doubly negatively charged $\CBVB$ pair is the
only defect that can reach concentrations
$10^{14}\mbox{--}10^{15}~\mbox{cm}^{-3}$ in bulk samples grown under
N-rich conditions. Our calculations show that these pairs do not have
an internal optical transition. Thus, they are unlikely to be SPEs
observed in Ref.~\cite{Mendelson2020}.

6. If oxygen is present during growth, it does not affect the
concentration of carbon defects under N-rich conditions. However,
under N-poor conditions, $\CNON$ defects are formed in densities
exceeding those of any other carbon- or oxygen-containing defects by
more than one order of magnitude. $\CNON$ defect is expected to emit
in the ultraviolet spectral region. It has been proposed as an
alternative defect to the carbon dimer to explain the 4.1~eV
luminescence~\cite{vokhmintsev2019}. Also, the incorporation of the
oxygen donor $\ON$ under N-poor conditions is significant, and this
pushes the Fermi level up. As a result, the concentration of odd
carbon clusters with surplus $\CN$ units is boosted compared to
oxygen-free conditions.

7. All other carbon defects not discussed above occur at substantially
smaller concentrations. This includes the $\CBVN$ defect, which was
recently proposed to be the origin of single-photon emission in
hBN~\cite{sajid2020} in the energy range of 1.6--2.1~eV. In this work,
we do not provide evidence for or against this attribution.  It is
firmly established that there exist several distinct classes of SPEs in hBN,
emitting in this energy range
\cite{Tran2016,Grosso2017,Exarhos2017}. However, our work provides
strong proof that $\CBVN$ cannot form in dense ensembles as observed in
Ref.~\cite{Mendelson2020}. If $\CBVN$ indeed emits in the 1.6--2.1~eV
energy range, thermodynamic modeling shows that there are emitters
that occur in much larger concentrations.

8. Overall, our results show that it is possible that the carbon
emitters observed in Ref.~\cite{Mendelson2020} are related to (i)
carbon monomers, (ii) carbon trimers, or (iii) star-like defects. The
conclusion about monomers reinforces the conclusion recently reached
by Auburger and Gali~\cite{Auburger2021}.  The idea of carbon trimers
as SPEs in the energy range 1.6--2.1~eV was first suggested by Jara
\emph{et al.}~\cite{Jara2021}.  Lastly, star-like defects have been
studied previously~\cite{Berseneva2011}, but their optical signal was hitherto
not discussed and they have not yet been proposed as visible SPEs. 
These defects, at variance with other candidates, have a high-spin ground state, 
and are thus interesting systems for further studies.

Table \ref{Table1} summarizes our main findings regarding the most
prominent carbon defects under both N-rich and N-poor
conditions. Defect densities corresponds to the situation without
oxygen present.

Stepping aside from the relevance of our work to SPEs and looking at
the materials chemistry of hBN more generally, our work indicates that
simultaneous presence of oxygen and carbon during growth has a
particularly detrimental effect on the material quality, especially
under N-poor conditions. This is due to the very low formation energy
of the $\CNON$ defect, which occurs in very large concentrations under
N-poor conditions. If the presence of oxygen and carbon cannot be
avoided, growth under more nitrogen-rich conditions is a pathway
to improve the quality of hexagonal boron nitride.

\section*{Acknowledgments}

We thank I. Aharonovich and S. Lany for useful comments. This work has received funding
from the European Social Fund (Project No.~09.3.3-LMT-K-712-23-0110) under
a grant agreement with the Research Council of
Lithuania. Computational resources were provided by the
Interdisciplinary Center for Mathematical and Computational Modelling
(ICM), University of Warsaw (Grant No.~GB81-6), and the High Performance
Computing center ``HPC Saul\.etekis'' in the Faculty of Physics,
Vilnius University.

\bibliography{carbon} 

%\includepdf[pages=-]{./SM/SM.pdf}

%} 

\end{document}

\title{Thermodynamics of carbon point defects in hexagonal boron nitride: Supplemental Material}

\author{Marek Maciaszek}
\author{Lukas Razinkovas}
\author{Audrius Alkauskas}

\noaffiliation
\maketitle

\begin{figure*}
\includegraphics[width=1\linewidth]{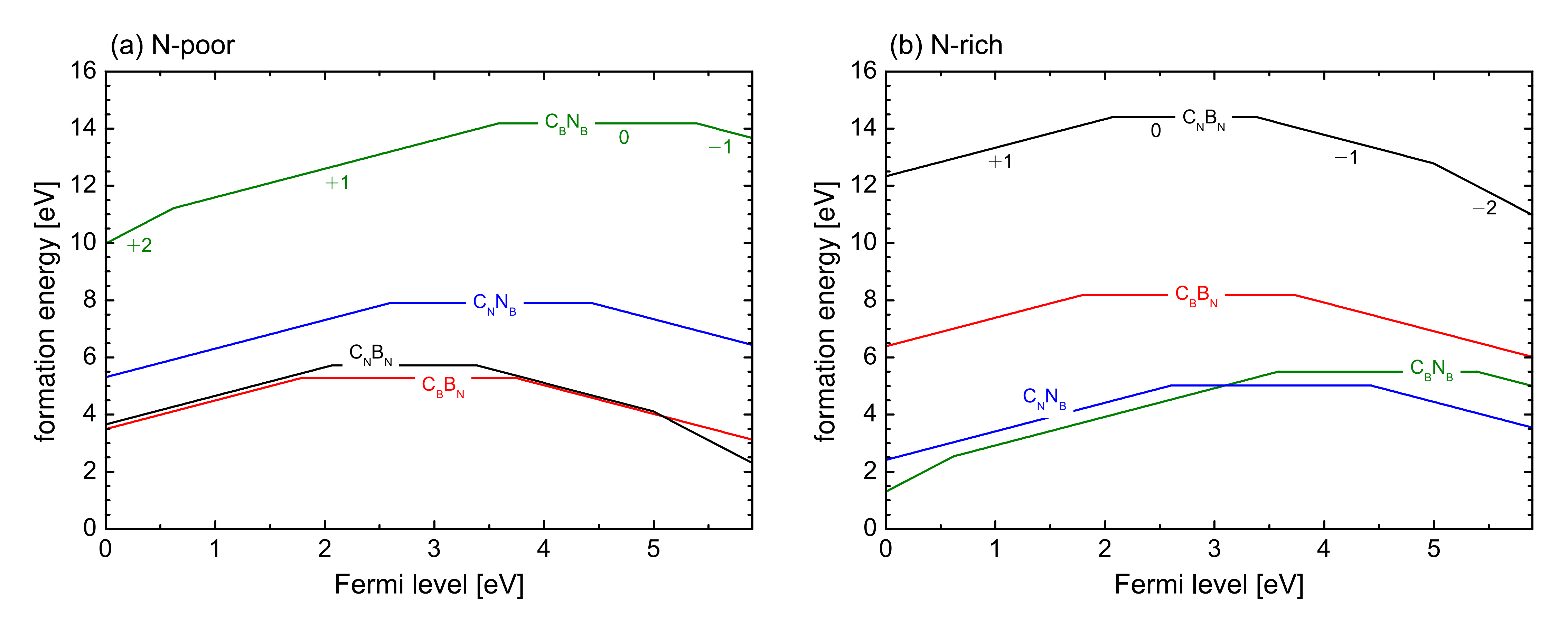}
\caption{Calculated formation energies of complexes of
  substitutional carbon with antisites as a function of the electron
  chemical potential: (a) N-poor, (b) N-rich conditions.}
\label{fig:antisites}
\end{figure*}

\section{Formation energies of complexes of substitutional carbon with
  antisites and vacancies}

\begin{figure*}
\includegraphics[width=1\linewidth]{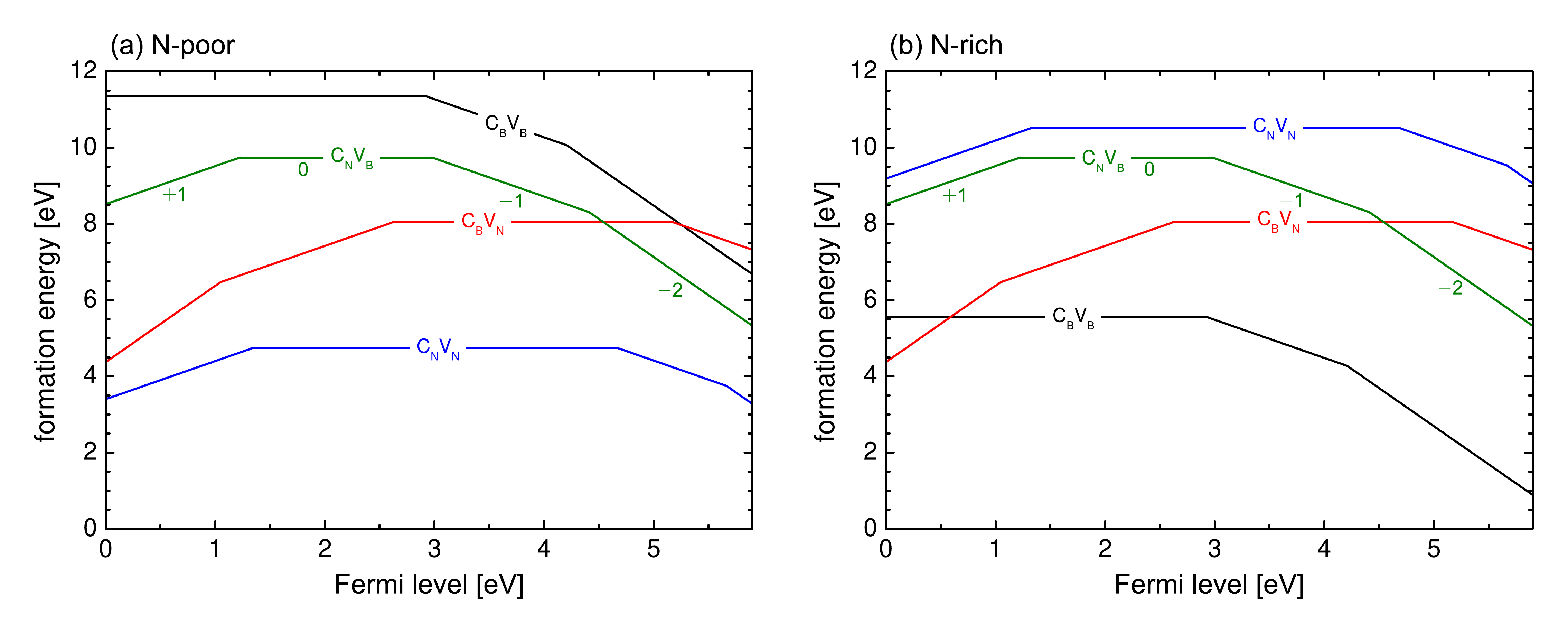}
\caption{Calculated formation energies of complexes of
  substitutional carbon with vacancies as a function of the electron
  chemical potential: (a) N-poor, (b) N-rich conditions.}
\label{fig:vacancies}
\end{figure*}

Calculated formation energies of complexes of substitutional
carbon with antisites are shown in Fig.~\ref{fig:antisites} for both
N-rich and N-poor conditions. Calculated formation energies of
complexes of substitutional carbon with vacancies are shown in
Fig.~\ref{fig:vacancies}.
\section{Formation energies of carbon pairs}

\begin{figure*}
  \centering
\includegraphics[width=0.5\linewidth]{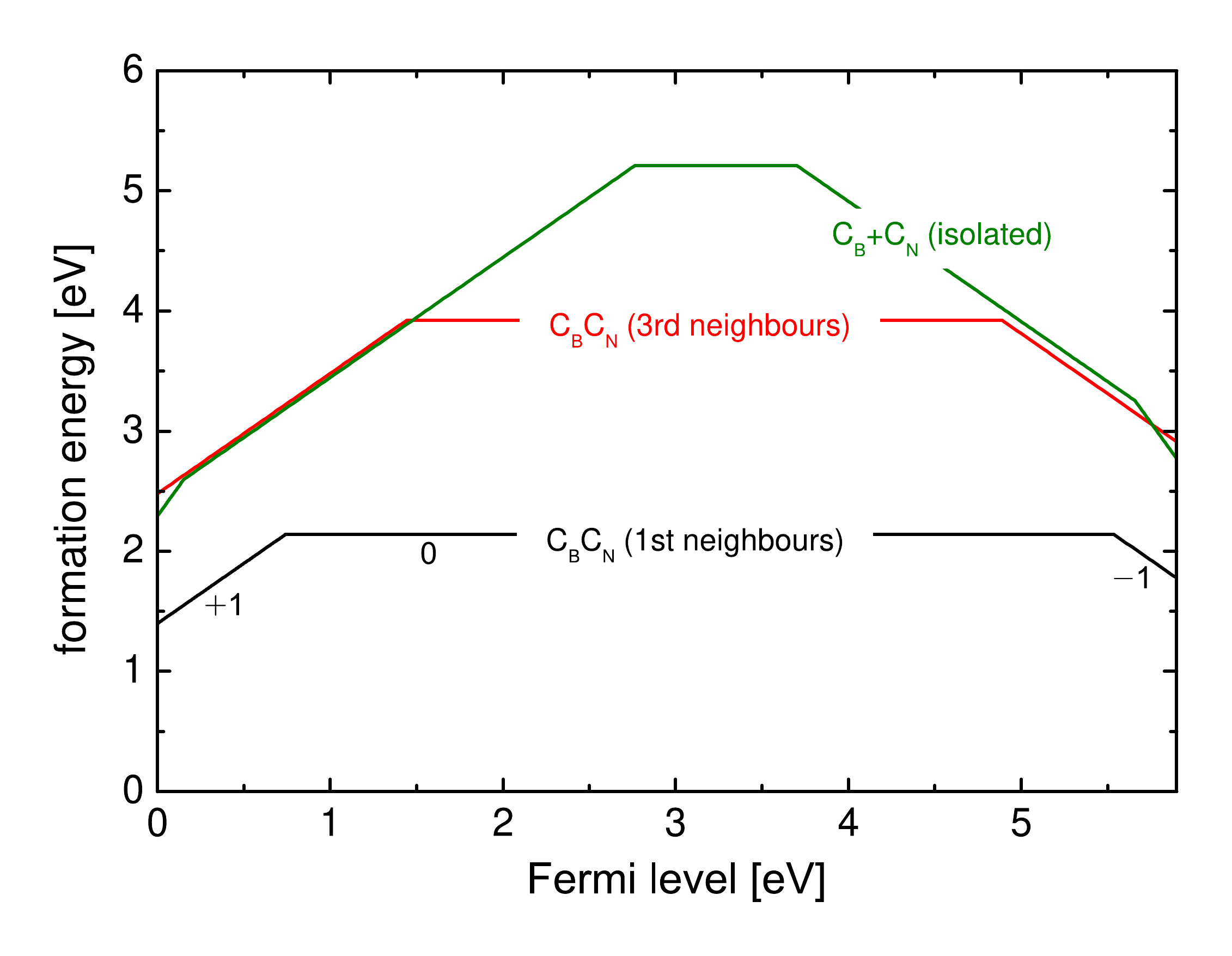}
\caption{Calculated formation energy of the 3rd-neighbor
  C$_{\text{B}}$C$_{\text{N}}$ pair as a function of the electron
  chemical potential. The formation energy of the nearest-neighbor
  carbon dimer C$_2$ (from Fig.~3 of the main text) and the formation
  energy of a $\text{C}_{\text{B}}\mbox{--}\text{C}_{\text{N}}$ pair at
  infinite separation (obtained from the results presented in Fig.~2
  of the main text) is shown for comparison.}
\label{fig:pairs}
\end{figure*}

The calculated formation energy of the 3rd-neighbor
C$_{\text{B}}$C$_{\text{N}}$ pair is shown in
Fig.~\ref{fig:pairs}. The formation energy is compared to that of the
nearest-neighbor carbon dimer C$_2$ (from Fig.~3 of the main text),
and the formation energy of a C$_{\text{B}}$--C$_{\text{N}}$ pair at
infinite separation (obtained from the results presented in Fig.~2 of
the main text). According to our simulations, the 3rd-neighbor
C$_{\text{B}}$--C$_{\text{N}}$ pair, as well as
C$_{\text{B}}$--C$_{\text{N}}$ pairs separated by larger distances,
occur at concentrations smaller than $N_D=10^{14}$ cm$^{-3}$ and are
thus not discussed in the main text.

\section{Formation of carbon point defects for a fixed concentration
  of carbon atoms}

As discussed in Sec. VI of the main text, the application of the
formation energy formalism implies fixed chemical potentials of atoms
during growth. This is a natural assumption in epitaxial growth, which
corresponds to the consideration of the grand canonical
ensemble~\cite{Zhang1991}. However, often it is more natural to
consider growth in the presence of a fixed amount of atoms rather than
fixed chemical potentials (e.g., during bulk growth). This is the
limit of the canonical ensemble. The formation energy formalism can be
still applied, but atomic chemical potentials are not input
parameters, but are rather determined by the actual number of atomic
species. In this Section we illustrate the procedure to determine
these atomic chemical potentials.

For simplicity, let us analyze hBN with no oxygen. The starting point
is a system of $N_{\text{B}}$ boron, $N_{\text{N}}$ nitrogen, and $N_{\text{C}}$ carbon atoms per
unit volume. As in the main text, $N_{D,q}$ is the density of defects
$D$ in charge state $q$, and $N_D = \sum_q N_{D,q}$ is the total
density of defects $D$. $n_\text{B}$, $n_\text{N}$, and $n_\text{C}$
is the number of extra atoms of boron, nitrogen, and carbon, needed to
form the defect $D$. We have the following equations for the densities
of boron, nitrogen, and carbon atoms:
\begin{align*}  
  & N_{\text{B}} = \frac{1}{2} N_{\text{sites}}+ \sum_{D} n_{\text{B}} N_{D},
  \\
  & N_{\text{N}} = \frac{1}{2} N_{\text{sites}}+ \sum_{D} n_{\text{N}} N_{D},
  \\    
  & N_{\text{C}} = \sum_{D} n_{\text{C}} N_{D}.
\end{align*}

We define the stoichiometry parameter $\delta$ as $\delta = N_{\text{B}}/N_{\text{N}}$.
The problem of defect formation for fixed amount of atomic species has
been considered in literature previously, e.g., in
Ref.~\cite{Windl2021}.

\begin{figure*}
\includegraphics[width=0.5\linewidth]{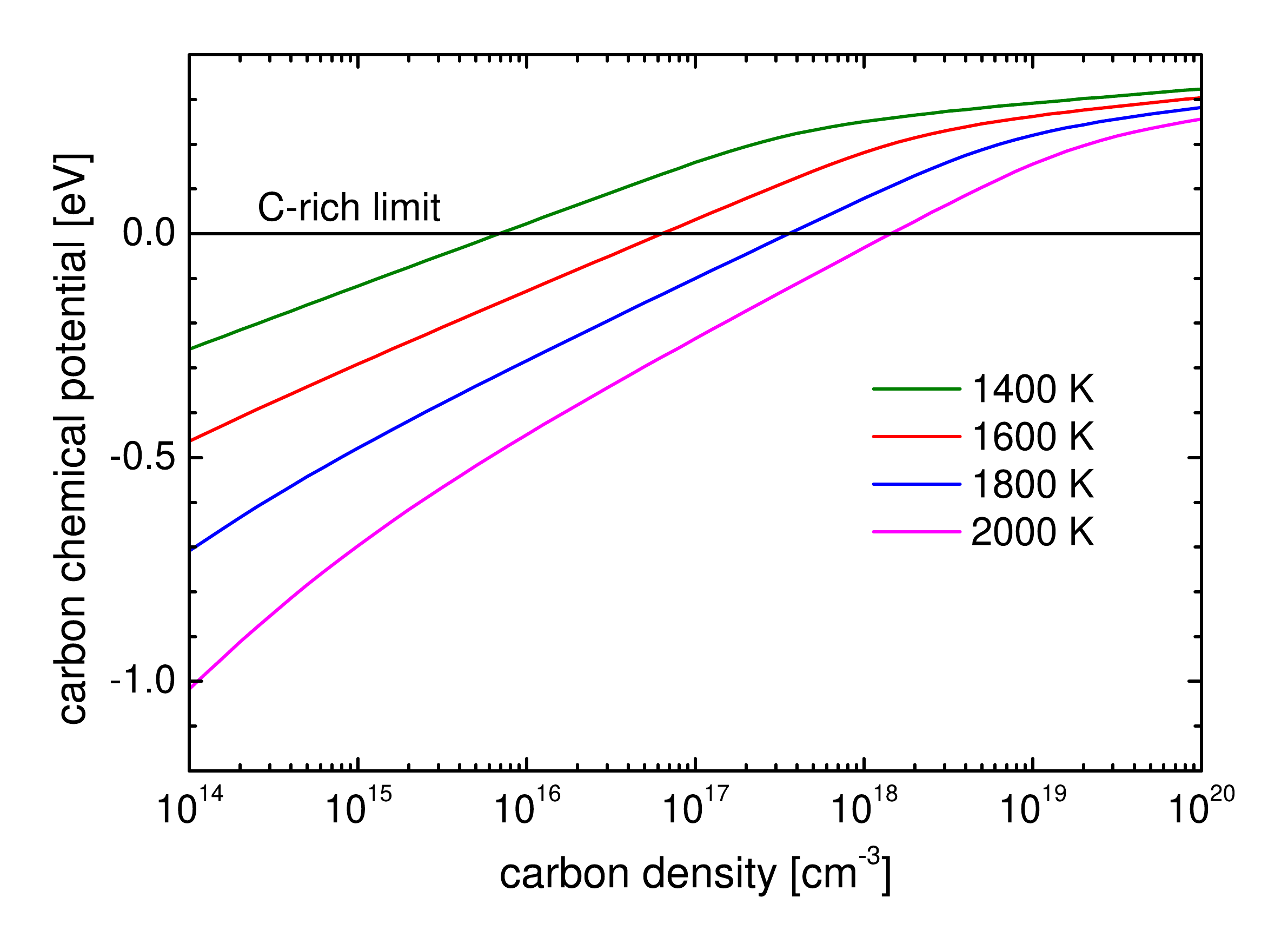}
\caption{Carbon chemical potential $\mu_{\text{C}}$ as a function of
  the carbon density $N_{\text{C}}$ for several different
  temperatures. The values of $\mu_{\text{C}}$ are referred to the
  energy of C atoms in graphite.  Only the values below the
  carbon-rich limit ($\mu_{\text{C}}=0$, black horizontal line) have
  physical meaning.\label{fig:miC}}
\end{figure*}

The procedure to determine atomic chemical potentials and the
concentration of defects was outlined in Sec. V of the main
text. Here, we illustrate this procedure for perfectly stoichiometric
hBN ($\delta=1$), focusing on the determination of the actual carbon
chemical potential. In Fig.~\ref{fig:miC} we show the calculated
values of carbon chemical potential $\mu_{\text{C}}$ as a function of
carbon concentration $N_{\text{C}}$ for a few different temperatures.
$\mu_{\text{C}}$ is referred to the energy of carbon in graphite.  As
$N_{\text{C}}$ increases, so does the chemical potential
$\mu_{\text{C}}$, reaching the carbon-rich limit ($\mu_{\text{C}}=0$,
black solid line) for a specific temperature-dependent carbon
concentration. Above the carbon-rich limit, the depicted curves of
$\mu_{\text{C}}$ have no physical meaning, and the real value of
$\mu_{\text{C}}$ is that corresponding to the black solid line in
Fig.~\ref{fig:miC}. When the carbon-rich limit is reached,
condensation of carbon point defects into graphite/graphene-like
carbon islands occurs. For the carbon densities considered in
Fig.~\ref{fig:miC}, we find that $\mu_{\text{B}}$ and $\mu_{\text{N}}$
are nearly independent of temperature and carbon concentration.

Once $\mu_{\text{C}}$ is determined, the actual concentration of
defects can be calculated using the methodology as described in
Sec.~II of the main text.

\bibliography{carbon}